\pdfoutput=1
\documentclass[ALICE,manyauthors]{cernphprep}
\usepackage[comma,square,numbers,sort&compress]{natbib}

\usepackage[T1]{fontenc} 
\usepackage[utf8]{inputenc}
\usepackage[english]{babel}

\usepackage{amssymb}
\usepackage{hyperref}
\usepackage{lineno}
\usepackage{subfigure}
\usepackage[amssymb]{SIunits}
\usepackage{booktabs}
\usepackage{multirow}
\usepackage{color}
\usepackage{tikz}
\DeclareGraphicsExtensions{.pdf,.png,.jpg}

\newcommand{\sqrts}{\sqrt{s}}

\newcommand{\TeV}{\mathrm{TeV}}

\newcommand{\gevc}{\mathrm{GeV}/c}
\newcommand{\tev}{\mathrm{TeV}}

\newcommand{\pt}{p_{\rm T}}

\newcommand{\Rt}{\ensuremath{R_{\rm T}}\xspace}
\newcommand{\Rtm}{\ensuremath{R_{\rm T}^{\rm meas}}\xspace}
\newcommand{\Rtt}{\ensuremath{R_{\rm T}^{\rm truth}}\xspace}
\newcommand{\Nch}{\ensuremath{N_{\rm ch}}\xspace}

\DeclareGraphicsExtensions{.pdf,.png,.jpg}

\definecolor{myblue}{rgb}{0.0,0.0,0.6}
\definecolor{darkred}{rgb}{0.7,0.0,0.0}
\definecolor{darkgreen}{rgb}{0,0.7,0.0}

\newcommand{\pT}      {\pt}
\newcommand{\ptlead}      {\ensuremath{p_{\mathrm{T}}^{\mathrm{leading}}}}

\newcommand{\avpt}    {\ensuremath{ \langle p_{\mathrm{T}} \rangle}}

\newcommand{\gmom}    {\mbox{${\rm GeV}/\mathrm{c}$}}
\newcommand{\mmom}    {\mbox{${\rm MeV}/\mathrm{c}$}}
\newcommand*{\degr}{\ensuremath{^\circ}}

\newcommand{\bfigFullPage}{\begin{figure} \begin{center} \vspace{0pt}}
\newcommand{\bfig}[1][!]{\begin{figure}[#1] \begin{center}}
\newcommand{\efig}{\end{center} \rule{4cm}{0.4pt} \end{figure}}
\newcommand{\bfigonecol}[1][!]{\begin{figure*}[#1] \begin{center}}
\newcommand{\efigonecol}{\end{center} \end{figure*}}
\newcommand{\efigNoLine}{\end{center} \end{figure}}
\newcommand{\btab}[1][!]{\begin{table}[#1] \begin{center}}
\newcommand{\etab}{\end{center} \rule{4cm}{0.4pt} \end{table}}
\newcommand{\btabonecol}[1][!]{\begin{table*}[#1] \begin{center}}
\newcommand{\etabonecol}{\end{center} \end{table*}}
\newcommand{\etabNoLine}{\end{center} \end{table}}

\newcommand{\bq}{\begin{equation}}
\newcommand{\eq}{\end{equation}}

\newcommand{\bqq}{\begin{eqnarray}}
\newcommand{\eqq}{\end{eqnarray}}

\newcommand{\ptmin}      {\ensuremath{p_{\mathrm{T}}^{\mathrm{track}}}}

\DeclareMathAlphabet{\mathpzc}{OT1}{pzc}{m}{it} 

\begin{document}%

\begin{titlepage}
\PHyear{2019}
\PHnumber{235}      
\PHdate{25 October}  

%

\title{Underlying Event properties in pp collisions at $\sqrt{s}$ = 13 TeV}
\ShortTitle{Underlying event in pp collisions at $\sqrts=13~\tev$}   
%
\Collaboration{ALICE Collaboration%
         \thanks{See Appendix~\ref{app:collab} for the list of collaboration 
                      members}}
\ShortAuthor{ALICE Collaboration}      
\begin{abstract}

This article reports measurements characterizing the Underlying Event (UE) associated with hard scatterings at midrapidity  ($ |\eta| < 0.8 $) in pp collisions at $\sqrts=13~\tev$. 
The hard scatterings are identified by the leading particle, the charged particle with the highest transverse momentum ($\ptlead$) in the event. 
Charged-particle number-densities and summed transverse-momentum densities are measured in different azimuthal regions defined with
respect to the leading particle direction: Toward, Transverse, and Away.
The Toward and Away regions contain the fragmentation products of the hard scatterings in addition to the UE contribution, whereas particles in the Transverse region are expected to originate predominantly from the UE.
The study is performed  as a function of $\ptlead$ with three different $\pt$ thresholds for the associated particles, $\ptmin >$ 0.15, 0.5, and 1.0 $\gevc$. 
The  charged-particle density in the Transverse region rises steeply for  low values of \ptlead and reaches a plateau.  The results confirm the trend 
that 
the charged-particle density in the Transverse region shows a stronger increase with $\sqrts$ than the inclusive charged-particle density at midrapidity. The UE activity is increased by approximately 20\% when going from 7 $\tev$ to 13 $\tev$ pp collisions.
The plateau in the Transverse region ($5 < \ptlead < ~ 40 \; \gmom$) is further characterized by the probability distribution of its charged-particle multiplicity normalized to its average value (relative transverse activity, $\Rt$) and the mean transverse momentum as a function of $\Rt$. 
Experimental results are compared to model calculations using PYTHIA~8 and EPOS~LHC. The overall agreement between models and data is within 30\%. These measurements provide new insights on the interplay between hard scatterings and the associated UE in pp collisions.
  \end{abstract}
\end{titlepage}
\setcounter{page}{2}

\section{Introduction}
\label{sec:intro}
In proton--proton (pp) collisions, particles originating from partonic scatterings with large 4-momentum transfer $Q$ compared to the Quantum Chromodynamics (QCD) scale $\Lambda_{\rm QCD}$, hard processes, are 
accompanied by additional, predominantly low transverse momentum ($\pt$), 
particles from the proton break-up (beam remnants) and possibly 
further scatterings, termed Multiple Parton Interactions (MPI)~\cite{MPIbook}. 
This associated particle production represents an important background to most  observables at hadron colliders and its detailed 
understanding and modeling with Monte Carlo (MC) generators is crucial for precision measurements and for connecting experimental observables to theory.
The empirical models for the description of the non-perturbative aspects in a high-energy scattering event evolution do not allow to clearly separate
particles originating from hard processes and the associated event activity event-by-event. In order to enable experimental studies and model comparisons one commonly separates the kinematic region containing the direct fragmentation products of the partons produced in the hardest scattering from the remaining part, generally referred to as the  Underlying Event (UE). The UE also contains particles from initial- and final-state radiation related to the hard interaction.

The first study of this kind was performed by the UA1 experiment at CERN's proton-antiproton ($\mathrm{Sp\bar{p}S}$) collider
by measuring the transverse energy density  
outside the leading jet cone~\cite{PLB132_1983_214,ua1_underlying_1,ua1_underlying_2}, the so-called jet pedestal region.
In the method introduced by CDF~\cite{cdf1} and used in the present analysis, one identifies the leading jet, or any other leading object in the event, and measures particle production in the azimuthal region orthogonal to the direction of this leading object, the Transverse region. Based on this method, several UE studies at the Tevatron~\cite{cdf1,cdf2,cdf3,cdf4} and at the LHC~\cite{EPJC70_2010_555,PRD83_2011_112001,A51_ALICE_UE,Aad2014,JHEP03_2017_157}, at various center-of-mass energies ($\sqrt{s}$), have been published.
These also include UE measurements in Drell-Yan~\cite{Chatrchyan:2012tb} and Z-boson~\cite{EPJC74_2014_3195,Sirunyan:2017vio,Aad2019} events performed by CMS and ATLAS.

A common characteristic of UE measurements at all collision energies is that the particle density in the Transverse region as a function of the $\pt$ of the leading object ($\ptlead$) rises steeply at low $\ptlead$ until a plateau at about twice the inclusive particle density is reached~\cite{PLB132_1983_214}.
In the framework of MPI-based models, the probability for a
hard scattering increases with the matter overlap in the collisions (decreasing pp impact parameter). And conversely, requiring a high-$\pt$ object to be detected in a given collision biases the selection of collisions towards those with a smaller impact parameter, at which the probability for additional uncorrelated scatterings and consequently the charged-particle number-density is enhanced~\cite{mpi_impact_param}. 
The charged-particle number-density (${\rm d}N_{\rm ch}/{\rm d}\eta$) in the plateau region increases logarithmically with $\sqrt{s}$ and faster than in minimum-bias events~\cite{A51_ALICE_UE}.
In MPI-based models, the height of the plateau is sensitive to the pp impact parameter dependence of the number of MPI per event~\cite{mpi_impact_param}. Hence, UE measurements have facilitated the implementation and tuning of such models~\cite{mpi_impact_param, Sjostrand:2004pf, Sjostrand:2004ef, Khachatryan:2015pea}. They have been used as tools for high precision 
Standard Model (SM) measurements as well as searches for physics beyond the SM.
In recent years it has been shown that they are also important to obtain a qualitative 
understanding of the centrality dependence of hard processes in p--Pb~\cite{Adam:2014qja} and Pb--Pb 
\cite{Morsch:2017brb, Acharya:2018njl} collisions at LHC energies.

During the last decade, the study of the bulk properties of pp collisions has gained increased 
interest as a research field in its own right. One of the most important discoveries in pp 
collisions at the LHC is the observation of collective, fluid-like features.
They are strikingly similar to those observed in heavy-ion collisions (AA), where they are attributed to 
the production of a deconfined hot and dense medium, known as the Quark-Gluon Plasma
(see Ref.~\cite{Nagle:2018nvi} for a recent review). 
The question arises whether the conditions created in high-multiplicity pp collisions can also 
modify, as in AA~\cite{Norbeck:2014,Armesto:2015ioy}, 
the yields of hard probes, for example through partonic energy loss~\cite{Zakharov:2014mda}.
Hence, the study of hard processes as a function of the charged-particle number-density has moved 
into the focus of interest. In this context the UE activity in the Transverse 
region (particle or summed-$\pt$ density) provides an event-activity classifier 
with reduced sensitivity to the hard process studied, which 
compared to inclusive classifiers can reduce trivial auto-correlation effects~\cite{Martin:2016igp, Weber:2018ddv}. With this in mind, the measurement of the distribution of the number-density in the Transverse 
region normalized to its average (relative transverse activity, $\Rt$)~\cite{Martin:2016igp} is included in the set of UE measurements reported in this paper.

This paper reports measurements characterizing the UE associated with hard scatterings performed at midrapidity  ($|\eta| < 0.8$) in pp collisions at
$\sqrts=13~\tev$ based on the CDF method~\cite{cdf1}, which utilizes the 
leading-charged particles. 
It extends the previous measurement of the number-density and summed-$\pt$ densities using charged particles~\cite{JHEP03_2017_157} to a lower $\pt$
threshold, $\ptmin >$ $0.15 \; \gmom$, in order to get higher sensitivity to the soft part of the UE.
The results are compared to the previous ALICE measurements in the same kinematic regions for pp collisions at $\sqrt{s} = 0.9$ and $7 \; \tev$.
The plateau in the Transverse region ($5 < \ptlead < ~ 40 \; \gmom$) is further characterized by the probability distribution of its charged-particle number-density normalized to its average value ($\Rt$). Moreover, the mean-transverse momentum in the Transverse region is studied as a function of $\Rt$. 

The paper is organized as follows: Section~\ref{sec:ue} introduces the UE observables. 
The MC event generators used in this paper are described in Sec.~\ref{sec:mc}.
The ALICE subsystems used in the analysis are described in Sec.~\ref{sec:detector}, 
and Sec.~\ref{sec:evsel} is dedicated to the analysis and data correction procedures, which includes 
the evaluation of the systematic uncertainties. 
The final results are presented and discussed in Sec.~\ref{sec:results} and the conclusions are summarized in
Sec.~\ref{sec:conclusion}.

\section{Underlying Event observables}
\label{sec:ue}
\begin{figure*}[htbp]
  \begin{center}
     \begin{tikzpicture}[>=stealth, very thick, scale=1.1]
       \small

       \draw[color=blue!80!black] (0, 0) circle (3.3);
       \draw[rotate= 30, color=gray] (-3.3, 0) -- (3.3, 0);
       \draw[rotate=-30, color=gray] (-3.3, 0) -- (3.3, 0);

       \draw[->, color=black, rotate=-2] (0, 3.5) arc (90:47:3.5) node[right] {$\Delta{\varphi}$};
       \draw[->, color=black, rotate=2] (0, 3.5) arc (90:133:3.5) node[left] {$-\Delta{\varphi}$};

       \draw[->, color=red, ultra thick] (0, 2) -- (0, 4.5) node[above,align=center,color=red] {Leading-particle};
       \draw[->, color=red!70!black, very thick, rotate around={  10:(0,0)}] (0, 2) -- ( 0.0, 3.8);
       \draw[->, color=red!70!black, very thick, rotate around={ -10:(0,0)}] (0, 2) -- ( 0.0, 3.8);
       \draw[->, color=blue!70!black, ultra thick] (0, -2) -- ( 0.0, -4);
       \draw[->, color=blue!70!black, ultra thick, rotate around={-15:(0,-0.3)}] (0, -2) -- ( 0.0, -3.6);
       \draw[->, color=blue!70!black, ultra thick, rotate around={ 15:(0,-0.3)}] (0, -2) -- ( 0.0, -3.6);
       \draw[->, color=green!70!black, very thick, rotate around={ 65:(0,0)}] (0, -2.6) -- ( 0.0, -3.6); 
       \draw[->, color=green!70!black, very thick, rotate around={ 105:(0,0)}] (0, -2.6) -- ( 0.0, -3.6); 
       \draw[->, color=green!70!black, very thick, rotate around={ -65:(0,0)}] (0, -2.6) -- ( 0.0, -3.6); 
       \draw[->, color=green!70!black, very thick, rotate around={ -105:(0,0)}] (0, -2.6) -- ( 0.0, -3.6); 
       \draw (0,  1.3) node[align=center] {Toward\\ $|\Delta\varphi| < 60^\circ$};
       \draw (0, -1.3) node[align=center] {Away\\ $|\Delta\varphi| > 120^\circ$};
       \draw (3.2, 0) node[align=center, anchor=east] {Transverse\\ $60^\circ < |\Delta\varphi| < 120^\circ$};
       \draw (-3.2, 0) node[align=center, anchor=west] {Transverse\\ $60^\circ < |\Delta\varphi| < 120^\circ$};
     \end{tikzpicture}
    \caption{Illustration of the Toward, Transverse, and Away regions in the azimuthal plane with respect to the leading particle direction.}
    \label{fig: uedef}
  \end{center}
\end{figure*}
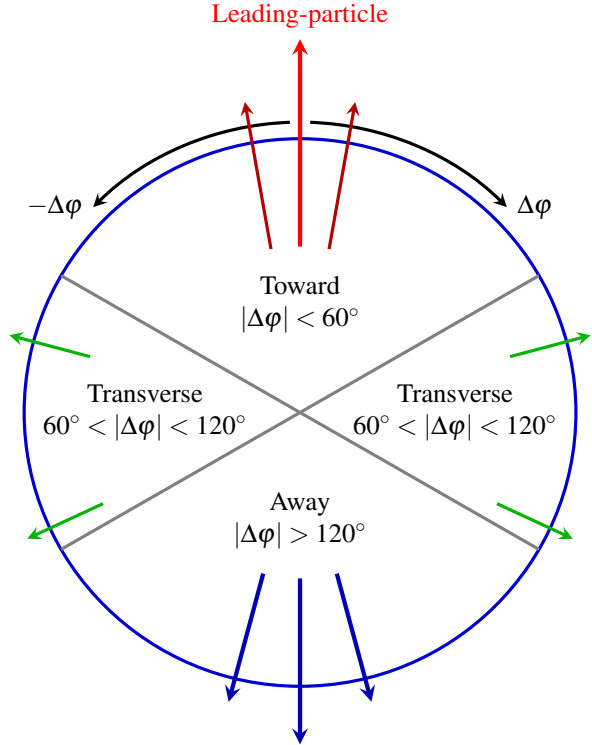

The UE observables considered in this study are based on primary charged particles\footnote{A primary particle is a particle with a mean proper lifetime $\tau$ larger than 1~cm/$c$, which is
either produced directly in the interaction, or from decays of particles with $\tau$ smaller than 1~cm/$c$, restricted to decay chains leading to the interaction~\cite{alicePrimary}.}
reconstructed
in the pseudorapidity range $|\eta| < 0.8$ with three different thresholds of the transverse momentum:
$\ptmin > $ 0.15, 0.5, and 1.0 {\gmom}, for both the leading particle ( \ptlead ) and the associated particles used in the correlation studies.
The UE observables are measured in three different 
regions defined by the relative azimuthal angle, $|\Delta\varphi| = \varphi - \varphi_{\rm leading}$, 
to the direction of the leading-charged particle (see Fig.~\ref{fig: uedef}):
\begin{itemize}
\item $|\Delta\varphi| < 60\degr$, the Toward region,
\item $60\degr < |\Delta\varphi| < 120\degr$, the Transverse region,
\item $|\Delta\varphi| > 120\degr$, the Away region.
\end{itemize}

The following observables, measured as a function of \ptlead, are considered to characterize the UE:
\begin{itemize}
\item average charged-particle density:
\begin{equation}  
\label{eq_numbdens}      
\frac{1}{\Delta\eta \times \Delta\varphi} \frac{1}{N_{\rm ev}(\ptlead)}N_{\rm ch}
\end{equation}
\item average summed-$\pT$ density:
\begin{equation} 
\label{eq_sumpt}       
\frac{1}{\Delta\eta \times \Delta\varphi} \frac{1}{N_{\rm ev}(\ptlead)}\sum \pT
\end{equation}
\end{itemize} 
evaluated in the three azimuthal regions, where $N_{\rm ev}(\ptlead )$ is the number of 
events satisfying a given  $\ptlead$ interval,   $\Delta \varphi$ = 2$\pi$/3 is
the  width of the regions in azimuth, and  $\Delta \eta =$~1.6 is  the acceptance window in pseudorapidity. 
The leading particle is not included in the calculation of the particle density and in the summed $\pt$ of the Toward region.

The  \ptlead\ can be regarded as a suitable proxy for the transverse-momentum scale of the hard 
scattering to avoid any problems related to jet reconstruction at low transverse momentum.
The restriction of the leading-particle pseudorapidity to the acceptance of the detector is part of the 
definition of the observables. In particular, the measurements did not correct for the fact that 
particles with $\pT > \ptlead$ can be present outside the acceptance. Therefore, the same selection is also applied in MC simulations.

The Toward and Away regions are predominated by particle production from the hard process and are, therefore, relatively insensitive to the softer UE. Conversely, the Transverse region is more sensitive to the UE as this 
region is least affected by contributions from the hardest scattering~\cite{cdf1}. 
Observables defined inside this region are the primary focus of UE measurements. 

For \ptlead~above the onset of the jet pedestal plateau, the UE depends only weakly on this quantity. It has therefore been proposed in~\cite{Martin:2016igp} to study the UE properties in events that contain one leading object with \ptlead\ in the range of the plateau, as functions of a new variable for quantifying event activity, relative transverse activity,~\Rt, defined as:   
\begin{equation}
R_{\rm T} = \frac{N_{\rm inc}}{\langle N_{\rm inc} \rangle},
  \label{eq: rt}
\end{equation}
where $N_{\rm inc}$ is the inclusive number of charged particles in an event 
and $\langle N_{\rm inc} \rangle $ is the event-averaged number-density, both evaluated in the Transverse region. 
Using this observable as an event classifier one can, as proposed in~\cite{Martin:2016igp},
test whether events with very small UE activity are compatible with equivalent measurements in $e^{+} e^{-}$ collisions (jet universality) or whether the scaling behaviour towards high UE activity exhibits properties of non-trivial soft-QCD dynamics.
As a self-normalized observable, \Rt\ is relatively insensitive to center-of-mass energy and kinematic selection variations, while simultaneously covering a large dynamic range in terms of event activity. 
The present paper reports the first measurement of the $\Rt$ probability distribution and the mean transverse momentum \avpt\ in the Transverse region as a function of $\Rt$.

\section{Monte Carlo models}
\label{sec:mc}
Particle production in hadronic collisions can be classified according to the energy scale of the process
involved. At high-momentum transfers, $Q^2\gg \Lambda_{\rm QCD}^2 $, perturbative Quantum Chromodynamics
(pQCD) is the appropriate theoretical framework to describe partonic interactions. 
This approach can be used to quantify parton yields and correlations, whereas the transition from partons to 
hadrons (hadronization) is a non-perturbative process that has to be treated using phenomenological 
approaches. 
For momenta of the order of the QCD scale, $\sim 200 \, \mmom$, a perturbative treatment is no longer feasible. Furthermore, at the center-of-mass energies of the LHC, with momentum transfers of a few \gmom, the calculated QCD cross sections for $2 \rightarrow 2$ parton scatterings exceed the total hadronic cross section for pp collisions (see for example~\cite{Sjostrand:2004pf}). 
This suggests that hard MPI occur in this regime. 
The overall event dynamics cannot be derived fully from first principles and must be described using phenomenological models implemented as general purpose MC generators. In these event generator implementations, model-specific choices are made to regulate these processes at low momentum scales.  This section reviews relevant features of the \textsc{Pythia}~8~\cite{Pythia8} and \textsc{Epos}~\cite{Drescher:2000ha} MC event generator models, which are used in this study for data correction and for comparison to our fully corrected results. 
A more detailed description of different general-purpose MC generators can be found for example in~\cite{PhysRevD.98.030001}.

\paragraph{\textsc{Pythia}~8}
In \textsc{Pythia}, event generation starts with a primary process that defines the nature of 
the event. At LHC energies, this is in most cases a leading order pQCD partonic scattering. 
At small $\pT$ values, color screening effects need to be taken into account. Therefore a cut off, $p_{\rm T,0}$, is introduced, which damps the QCD cross section for $\pT \ll p_{\rm T,0}$. This cut off is one of the main tunable model parameters.
Subsequent partonic processes calculable in pQCD are initial- and final-state radiation 
interleaved with MPI, and the structure of beam remnants. The number of MPI in this model
depends on the impact parameter of the pp collision. 
After these steps, a realistic partonic structure including 
jets and UE activity is obtained. The partonic configuration is hadronized 
using string fragmentation as described by the Lund string model~\cite{lund_string}, followed by the 
decays of unstable particles. In  collisions with several MPI, individual long strings 
connected to the remnants are replaced by shorter additional strings connecting partons from different MPIs. This mechanism, called 
color reconnection, has been introduced to reproduce the increase of the average transverse 
momentum with multiplicity observed in data~\cite{Sjostrand:2006za}.
For the comparison with measured observables, MC simulated samples with Monash-2013~\cite{Pythia8Monash} tune and NNPDF2.3 LO PDF set)~\cite{PDF_NNPDF} are used.

\paragraph{\textsc{Epos} LHC}
The \textsc{Epos}~\cite{Werner:2005jf} event generator can be used to simulate pp, pA, and AA collisions.
The multiple scattering approach in \textsc{Epos}  is based on a combination of Gribov-Regge theory 
and pQCD~\cite{Drescher:2000ha}. An elementary scattering corresponds to a parton ladder, containing a hard scattering calculable based on pQCD, including initial- and final-state radiation. Parton ladders which are formed in parallel to each other share the total collision energy leading to a consistent treatment of energy conservation in hadronic collisions.
String hadronization in \textsc{Epos} is based on the local density of string segments per unit volume with respect to a critical-density parameter. 
Event-by-event, string segments in low-density regions hadronize normally and independently, creating the corona, while string segments in high-density regions are used to create a core with collective expansion and hadronization including radial and longitudinal flow effects.
The \textsc{Epos} LHC tune considered here is based on a dedicated parameter set used to describe data from all LHC energies and collision systems~\cite{Pierog:2013ria}. 


\section{Experimental setup}
\label{sec:detector}
ALICE is the dedicated heavy-ion experiment at the LHC. A detailed description of the ALICE detectors can be found in~\cite{ALICEexp}. In the following, only the detector components used in the data analysis presented here are described. 

The ALICE apparatus comprises a central barrel (pseudorapidity coverage $|\eta|<0.9$ over full azimuth) situated in a uniform $0.5\ \mathrm{T}$ magnetic field along the beam axis ($z$) supplied by a large solenoid magnet. The forward and backward rapidity plastic scintillator counters, V0A and V0C, are positioned on each side of the interaction point, covering pseudorapidity ranges $2.8<\eta<5.1$ and $-3.7<\eta<-1.7$, respectively. And they are used for determination of the interaction trigger and to suppress beam-gas and beam-halo background events. The central barrel contains a set of tracking detectors: a six-layer high-resolution silicon Inner Tracking System (ITS) surrounding the beam pipe, and a large-volume ($5\ \mathrm{m}$ length, $0.85 \ \mathrm{m}$ inner radius and 2.8 m outer radius) cylindrical Time Projection Chamber (TPC). 

The first two layers of the ITS are equipped with high-granularity Silicon Pixel Detectors (SPD), which cover the pseudorapidity ranges $|\eta| < 2.0$ and $|\eta| < 1.4$ respectively. 
The position resolution is $12 \, {\rm \mu m}$ in r-$\varphi$ and about $100 \, {\rm \mu m}$ along the beam direction.
The following two layers are composed of Silicon Drift Detectors (SDD). The position along the beam direction is measured via collection anodes, and the associated position resolution is about $50 \, {\rm \mu m}$. 
The r-$\varphi$ coordinate is given by a drift-time measurement, with a spatial resolution of about $60 \, {\rm \mu m}$. 
Finally, the two outer layers are made of double-sided Silicon micro-Strip Detectors (SSD) with a position
resolution of $20 \, {\rm \mu m}$ in r-$\varphi$ and about $800 \, {\rm \mu m}$ along the beam direction.
The material budget of all six layers, including support and services, amounts to 7.7\% of a radiation length in the transverse plane.

The TPC covers the pseudorapidity range of about $|\eta| < 0.9$ for tracks traversing the outer radius.
In order to avoid border effects, the fiducial region has been restricted in this analysis to $|\eta| <  0.8$. 
The position resolution along the r-$\varphi$ coordinate varies from $1100 \, {\rm \mu m}$ 
at the inner radius to $800\, {\rm \mu m}$ at the outer. 
The resolution along the beam axis ranges from $1250\, {\rm \mu m}$ to $1100\, {\rm \mu m}$. 

The ITS and TPC space points are combined to reconstruct tracks from charged particles over a wide transverse momentum range starting from $p_\mathrm{T} = 0.15\  \mathrm{GeV}/c$. 
The tracking efficiency estimated from a full simulation of the detectors is $\approx 65\%$ at the lowest $\pT$, and increases with $\pT $, plateauing at  $\approx 80\% $ for $\pT > 2\, \gmom$. The  transverse momentum resolution is better than $3\%$ for primary tracks below $1\ \mathrm{GeV}/c$, and degrades linearly up to $6\%$ at $p_\mathrm{T}=40\ \mathrm{GeV}/c$ ~\cite{Abelev:2014ffa}. 
The transverse impact parameter resolution decreases from $300 \, {\rm \mu m}$ at $0.2 \, \gmom$ to  $20 \, {\rm \mu m}$ at $30 \, \gmom$.


\section{Analysis procedure}
\label{sec:evsel}
\subsection{Event selection}
\label{data}
The measurements presented here use data collected by ALICE during the 2016 LHC pp run at $\sqrt{s}=13~\,\mathrm{TeV}$. During this period, minimum-bias (MB) events were selected using the high purity V0-based MBand trigger which required a charged-particle signal coincidence in the V0A and V0C arrays. It is the same trigger as used in LHC Run 1 high-luminosity data taking~\cite{Abelev:2014ffa}. 
 After event selection, a data sample of 46.2 million MBand triggered events is obtained. 
Further event selection for offline analysis is made by requiring a primary vertex position along the $z$-axis within $\pm 10\ \mathrm{cm}$ ($|v_z| < 10$~cm) around the nominal interaction point to ensure full geometrical acceptance in the ITS. Pile-up interactions are limited by keeping the average number of pp interactions per bunch crossing below $0.06$ through beam separation in the horizontal plane. Residual pile-up events are rejected based on a multiple interaction points finding algorithm using SPD information~\cite{Abelev:2014ffa}. Primary tracks satisfying various quality selection criteria, described  in
the next section, are used in this analysis. Moreover, at least one track with a minimum transverse momentum  $\ptmin =$ 0.15, 0.5, and 1.0 $\gevc$ in the acceptance range $| \eta|<0.8$ is required for the analysis. 
%
%
Table~\ref{tab:events} summarizes the percentage of events remaining after each event selection step for the various $\ptmin$ selections. The last row in the table shows the sample size available for the \Rt analysis. 
\begin{table}[bhtp]
\small 
\caption{\normalsize Events (absolute numbers and percentages) remaining after each event selection step.}
\centering  
\begin{tabular}{ 
   m{3.45cm}             |  
   m{1.25cm}<{\centering}   
   m{2.4cm}<{\centering}   
   } 
 \hline 
	& \textbf{Events} &  \textbf{ Fraction (\%) } \\
	\hline
	Offline trigger & 46.2 M & 100.0 \\
	Reconstructed vertex & 42.8 M & 92.6 \\
	$\ptmin\ > 0.15~\gmom$  & 40.4 M & 87.4 \\
	$\ptmin\ > 0.5~\gmom$  & 34.5 M & 74.7 \\
	$\ptmin\ > 1.0~\gmom$  & 22.1 M & 47.8 \\
	$\ptlead\ > 5.0~\gmom$  & 0.43 M & 0.93 \\
    \hline 
\end{tabular}
\normalsize 
\label{tab:events}

\end{table}
\subsection{Track selection}
\label{event_sel}
The selected charged-particle tracks are required to have at least 70 TPC space points, the number of geometrically possible clusters which can be assigned to a track, and more than $60\%$ of the findable TPC space points, i.e. those that can be assigned to tracks based on geometrical criteria derived from track parameters. 
The track selection criteria are optimized for good momentum resolution and minimal contamination from secondary tracks. 
For this purpose a track must have at least 3 clusters in the ITS, one of which has to be in the first 3 layers.
The quality of the track fitting measured in terms of the $\chi^{2}$ per space point is required to be lower than 4 (each space point having 2 degrees of freedom). 
Moreover, to reject secondary tracks, 
the distance of closest approach (DCA) of the track to the primary vertex along the beam axis (DCA$_{Z}$) is required to be smaller than 2~cm.
In the transverse direction, the maximum allowed DCA$_{XY}$ corresponds to seven times the $\pT$-dependent DCA$_{XY}$ resolution.
%
%
\subsection{Corrections}
\label{corrections}
For efficiency and acceptance corrections, events are generated using \textsc{Pythia}~8 MC with the same tune as listed in the early section. They are subsequently transported through the software description of the ALICE apparatus using GEANT 3.21~\cite{A32_RefGeant3}. 
For particles crossing sensitive detector layers the detector response is simulated. 
The simulated events are reconstructed and analyzed using the same algorithms as used for the real data. The number of simulated events similar to the ones in data are used to determine the corrections.


The measured particle and $\sum \pT$ densities are corrected for tracking efficiency, contamination from secondary particles, and the finite vertex reconstruction efficiency. 
The particle with the highest $\pT$ in a collision may not be detected due to finite acceptance and efficiency of the detection apparatus, and a lower $\pT$ track enters the analysis instead. If the misidentified leading particle has a different $\pT$ but roughly the same direction as the true leading particle, this leads to a shift in \ptlead. On the other hand, if the misidentified leading particle has a significantly different direction than the true one, this will cause a rotation of the event topology and a bias on the UE observables. 
The data are corrected for these effects using a data-driven procedure described in detail in Ref.~\cite{A51_ALICE_UE}.

Table~\ref{tab:corr} summarizes the maximum effect of each correction on the measured final observable for $\ptmin > 0.15~\gmom$. For other $\ptmin$ thresholds the effects are
similar. The correction procedures follow the same approach as described in ~\cite{A51_ALICE_UE}. 
In the following only the corrections for \Rt-related distributions using unfolding and re-weighting methods will be detailed.

\newcommand{\specialcell}[2][l]{\begin{tabular}[#1]{@{}l@{}}#2\end{tabular}}  
\renewcommand{\arraystretch}{1.5}
\begin{table}[bhtp]
\caption{\normalsize Maximum effect of $\pt$-dependent corrections on measured particle and summed-$\pT$ densities with $\ptmin > 0.15~\gmom$ threshold. 
}
\small 
\centering  
\begin{tabular}{ 
   m{4.45cm}             |  
   m{2.75cm}<{\centering}   
   m{1.25cm}<{\centering}   
   } 
 \hline 
\textbf{Correction} &  \textbf{ charged-particle density } & \textbf{$\sum \pt$  density} \\
	\hline
Efficiency  & 30$\%$ & 25$\%$ \\\hline
Leading track misidentification & 7$\%$ & 7$\%$ \\\hline
Contamination & 3$\%$ & 2$\%$ \\\hline
Vertex reconstruction & 0.9$\%$ & 0.9$\%$ \\\hline
\end{tabular}
\normalsize 
\label{tab:corr}
\end{table}

Due to the finite momentum resolution and tracking efficiency of the detector, the measured \Rt\ probability 
distribution and the charged-particle \avpt\  distribution as a function of \Rt are distorted. This affects, in particular, the \Rt\ probability 
distribution which falls steeply at large \Rt .
For this reason an unfolding procedure is employed to correct for detector effects. 
For the measurement of the charged-particle \avpt\  vs. \Rt\ both quantities have to be corrected. For the full 
unfolding a 4-dimensional response matrix is needed. Considering that the charged-particle \avpt\ rises slowly as a function of $\Rt$, a re-weighting correction procedure is performed, as described in 
~\cite{PLB2010_693_53, PLB2013_727_371}. 
\subsubsection*{Unfolding}
\begin{figure*}[hbtp]
 \begin{center}
  \includegraphics[width= 0.45\textwidth]{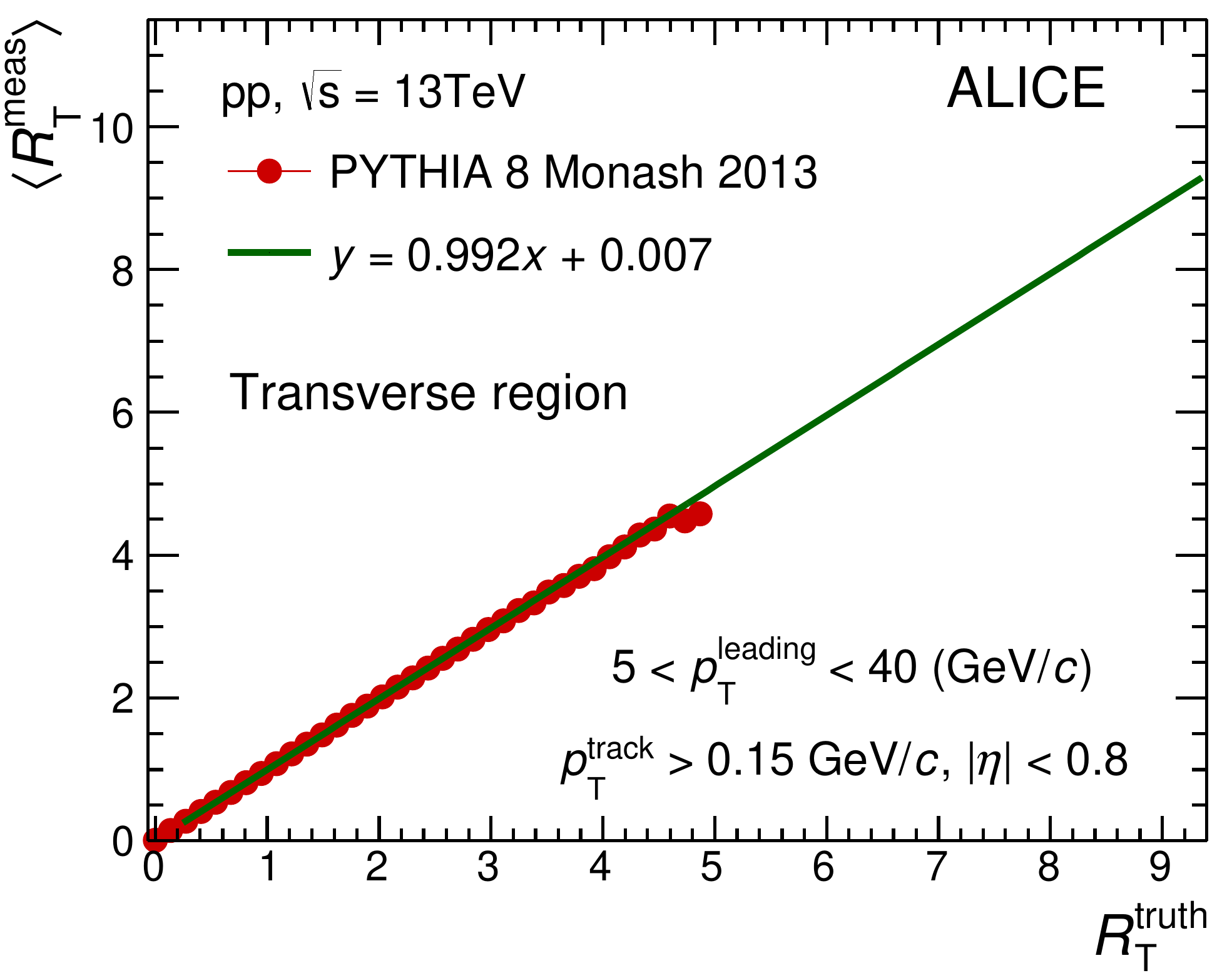}
  \includegraphics[width= 0.45\textwidth]{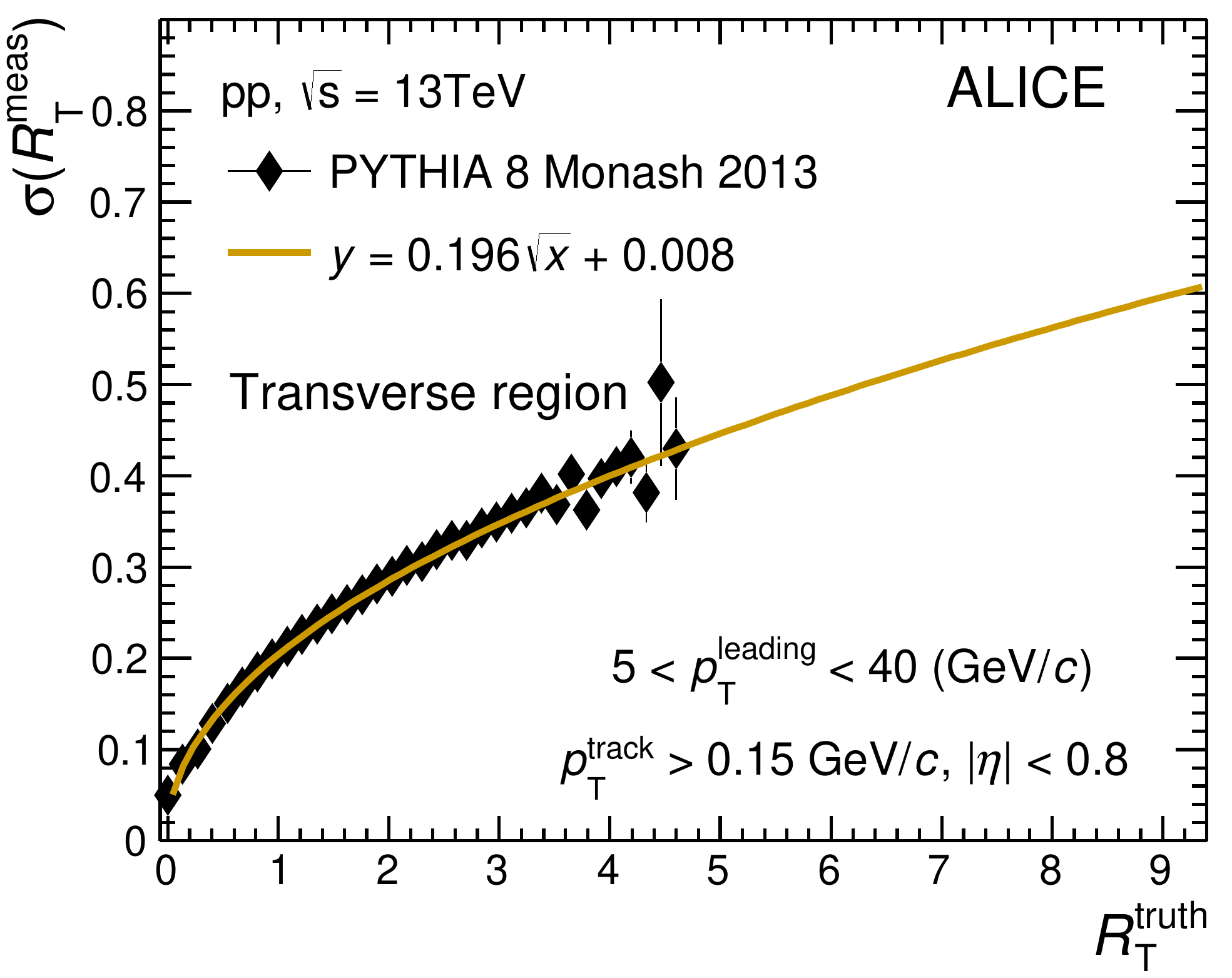}
 \end{center}
 \caption{
 Mean \Rtm\ and standard deviation $\sigma$ of the \Rtm\ distribution as a function of \Rtt .
 The solid lines represent the fits to the points and the extrapolation to higher \Rtt, resulting in the parameterized response matrix used for unfolding. 
 }
 \label{fig:RTMeanSigma}
 \end{figure*}

The \Rt\ probability distribution is corrected using 1-dimensional Bayesian unfolding~\cite{DAgostini:1994zf}, an iterative method based on Bayes' theorem, as implemented in the RooUnfold package~\cite{Adye:2011gm}. To this end, a 2-dimensional response matrix is 
created using the \textsc{Pythia}~8 generator. It maps the particle level \Rtt\ obtained from a MC simulation to the detector level \Rtm\ obtained 
after full GEANT3 based transport, track reconstruction, and track selection. 
Projections of the distribution of \Rtm\ for a given 
\Rtt\ are well approximated by a Gaussian distribution, and hence, described by its mean and standard 
deviation. This is used to extrapolate the response matrix into a region where statistical uncertainties 
are large and can deteriorate the quality of the unfolding.
Figure~\ref{fig:RTMeanSigma} shows the mean and standard deviation of \Rtm\ as a function of \Rtt\ obtained from the \textsc{Pythia}~8 simulation. A linear function and a function of the form $a\sqrt{x}+b$ are used to fit the mean and the standard deviation, respectively.
Over the whole $\Rtt$ range the relative difference between the fit and the simulated data is less than 1\% and less than 0.2\% for 
$\Rtt < 2$. 
Therefore, the results of the fits are used to extend the response matrix to  $\Rtt > 2$. The smoothened response matrix is used for \Rt probability distribution unfolding and \avpt\ re-weighting, with the difference between the data points and the parameterized fitting functions (as shown in Fig.~\ref{fig:RTMeanSigma}) being propagated to the final systematical uncertainties. 

For the unfolding correction, the \textsc{Pythia}~8 generated \Rtt\ distribution is used as the prior. Convergence is reached typically after three iterations. As an additional cross check, the analysis is also carried out using the Singular Value Decomposition (SVD) unfolding~\cite{A44_unfold-svd}. The relative difference between the SVD and Bayesian unfolded distributions is found to be below 2\%.
To study the influence of the corrections using a particular MC event generator, \textsc{Epos} LHC generator is also used to determine the unfolding response matrix and as prior in the unfolding process. The difference of the unfolded results using different MC generator corrections is considered as part of unfolding uncertainty.    



%
%


\subsubsection*{Mean $\pT$~re-weighting}
For the measurement of the charged-particle $\avpt$ as a function of \Rt , a re-weighting approach is used.
The procedure is implemented based on the following relation between the true and measured \Rt\:
\begin{equation}
\avpt  (\Rtt) = \sum P(\Rtm | \Rtt) \times \avpt (\Rtm)
  \label{eq: re-weight}
\end{equation}
where $P(\Rtm | \Rtt)$ is the normalised probability distribution of $\Rtm$ in a given \Rtt\ interval, which is obtained from the detector response matrix previously described. 

\subsection{Systematic uncertainties}
\label{systematics}
The evaluation of the systematic uncertainties on the charged-particle number density and $\sum \pt$ density follows closely the methods developed for inclusive charged-particle measurements~\cite{PLB2010_693_53, Abelev:2013ala, Adam:2015pza} and the UE measurements at lower collision energies~\cite{A51_ALICE_UE}.
Table~\ref{tab:systUE} summarizes the systematic uncertainties evaluated for the 
particle transverse momentum $\ptmin >$ 0.15 $\gevc$ threshold, for selected \ptlead~ranges. 
In the following, the individual sources of systematic uncertainty, listed in the first column of Tab.~\ref{tab:systUE}, will be described briefly. A detailed description of the procedures can be found in~\cite{A51_ALICE_UE}.
\begin{itemize}
    \item {\it ITS-TPC track matching efficiency:}
    Systematic uncertainties on the ITS and TPC detector efficiencies are estimated by comparing the experimental ITS-TPC track matching efficiencies with those obtained using the MC sample.
    \item{\it Track and vertex selection:}
    By applying the efficiency and contamination corrections, one accounts for those particles which are lost due to detector effects, vertex reconstruction inefficiency, and secondary tracks which have not been removed by the selection criteria. These corrections rely on detector simulations. Therefore, the systematic uncertainties were estimated by varying the choices of track parameter requirements and vertex reconstruction parameters.
    \item{\it Secondary particle contamination:}
     MC generators underestimate the production of strange particles in data. The effect on the secondary particle contamination correction was estimated by varying the strange particle fraction between the one given by PYTHIA and the one compatible with the 
     tails of the $DCA_{XY}$ distribution which are predominantly populated by secondaries.

    \item{\it Misidentification bias:}
    The uncertainty on the leading-track misidentification correction is estimated from the discrepancy between a data-driven correction used in the analysis and an alternative method based on simulations.
\item{\it MC non-closure:}
By correcting an MC generator prediction after full detector simulation with corrections extracted using the same generator, one expects to reproduce the input MC prediction within statistical uncertainty. This consideration holds true only if each correction is evaluated with respect to all the variables to which the given correction is sensitive. Any statistically significant difference between input and corrected
distributions is referred to as MC non-closure and is added in quadrature to the total systematic uncertainty.
\item{\it MC dependence}:
The difference in final distributions when applying corrections extracted using \textsc{Pythia}~ 8 or \textsc{Epos} LHC generators was quantified and added to the systematic uncertainty.
\end{itemize}
	\begin{table*}[htbp]
		\caption{Systematic uncertainties of the charged-particle number and summed-$\pt$ densities as a function of $\ptlead$ for the transverse momentum threshold $\ptmin\ > 0.15 \; \gmom$ in pp collisions at $\sqrt{s}= 13\; {\rm TeV} $. When more than one number is quoted, the values refer to the uncertainty in Toward, Transverse, and Away regions, respectively; they are independent of the azimuthal region in all other cases. 
		    }
	\label{tab:systUE}
	\begin{center}
	\footnotesize
	\begin{tabular}{l|c|c||c|c}
    & \multicolumn{2}{c||}{\textbf{Charged-particle density}} & \multicolumn{2}{c} {\textbf{Summed-$\pt$~density}} \\
     \hline
      &$\ptlead < 1 ~\gevc$ & $\ptlead > 6 ~\gevc$ & $\ptlead < 1 ~\gevc$ & $\ptlead > 6 ~\gevc$ \\ 
      \hline
    \textbf{ITS-TPC track matching} &  0.3\% &   2.3\%  &  0.4\% &   3.2\% \\
     \textbf{Track cuts} &  0.8\% & 1.6\%  & 0.7\% &   1.5\%  \\
     \textbf{Secondaries contamination} &  0.2\% &  0.2\% &  0.2\% &  0.2\% \\ 
      \textbf{Misidentification bias} &  0.7\% &  negligible  & 0.9\% &  negligible \\
    \textbf{Vertex reconstruction } &  0.3\% &  negligible & 0.3\% & negligible \\
      \textbf{MC non-closure }  &  0.9\%, 0.9\%, 1.1\% &  0.7\%,  0.1\%,  0.1\% & 0.4\%, 0.2\%, 0.4\% & 0.6\%, 0.3\%, 0.3\%\\
    \textbf{MC dependence } & 0.3\%,  0.6\%, 0.8\% &  1.7\%,  2.8\%,  2.8\% & 0.5\%,  0.5\%, 0.4\% & 1.0\%,  3.0\%, 3.0\%\\
     \hline
    \textbf{Total uncertainty } &  1.5\%,  1.5\%, 1.8\% & 3.4\%, 3.9\%, 3.9\% &  1.4\%, 1.4\%, 1.5\% &  3.9\%, 4.7\%, 4.6\%\\
      \hline
	\end{tabular}	
	\end{center}
	\end{table*}

Since \Rt\ is a self-normalized quantity evaluated event-by-event, the systematic uncertainties related to the ITS-TPC track matching efficiency and track cuts partially cancel each other.
A residual effect results from the fact that the $\pT$ spectrum gets harder with increasing multiplicity, leading to a difference between the $\pT$ weighted efficiencies for $N_{\rm inc} $ and $\langle N_{\rm inc} \rangle$, which is imperfectly accounted for through the MC-based response matrix. 
The resulting scale uncertainty on $\Rt$ has been estimated by varying the shape of the $\pT$ spectra within uncertainties taken from the spectrum analysis~\cite{Adam:2015pza}. The variations are propagated to the $\Rt$ distribution via the response matrix.

Another contribution to the systematic uncertainty on the \Rt\ distribution results from the event selection. 
This uncertainty is estimated by repeating the \Rt\ analysis using a more restrictive selection on the primary vertex, $|v_z| < 7$ cm. 
Since the vertex reconstruction efficiency increases with $N_{\rm inc} $, also the systematic uncertainty related to this efficiency cancels only partially in the ratio $N_{\rm inc} $ / $\langle N_{\rm inc} \rangle$ (in particular at low $N_{\rm inc} $, where the efficiency is low). This uncertainty is evaluated by changing the default selection from one minimum track contributing to the primary vertex to two tracks. The resulting difference of 2.7\% is assigned as the systematic uncertainty on the $N_{\rm inc} $ determination. 

To validate the unfolding procedure, and identify potential biases, closure tests are performed which 
compare the unfolded distribution to the particle-level truth in the MC simulation. Consistency 
of the unfolding procedure is also ensured by re-folding the solution to detector level and comparing it to
the uncorrected distribution used as input. The remaining difference of 0.3\% is assigned as uncertainty from MC non-closure. As discussed in the previous section, the Bayesian unfolding is employed as the default method. The number of iterations serves as a regularization parameter in Bayesian unfolding. Based on the closure test and convergence, four iterations were chosen as the default. To estimate the related systematic uncertainty, the iterations parameter is varied by $\pm 2$.  The unfolded results are quite stable against regularization parameter variations with a maximum deviation of 1.2\% at high-\Rt . 
As an independent cross-check, the SVD unfolding has also been used in the analysis and the difference between SVD and Bayesian unfolded results are found to be less than 0.4\%.
The corrected \Rt\ distribution is obtained after the unfolding,  which is performed using the detector response matrix computed based on simulations with \textsc{Pythia}~8 tune Monash-2013. 
This particular choice of MC event generator affects the prior used for unfolding. To investigate the systematic uncertainty from this particular choice, the prior distributions are varied using the deviations of fitted Negative Binomial Distribution (NBD) distribution (see Sec.~\ref{sec:results}). 



For the  measurement of the \avpt\ distribution vs. \Rt, a $1\%$ systematic uncertainty from MC non-closure using the re-weighting procedure is assigned. The uncertainty from the ITS-TPC track matching efficiency contributes to 1.3\%, while the track selection criteria results in a total uncertainty of around 0.4\%. The residual scale uncertainty due to the particle $\pT$-spectrum slope changes is estimated using the same approach as described for \Rt. Here both \Rt and \avpt\ are affected and therefore the residual scale uncertainty largely cancels out, resulting in a systematic uncertainty of $0.2\%$ on \avpt. The variations of the number of vertex contributors and the vertex cut contribute both 0.2 \% to the systematic uncertainty.

The maximum systematic uncertainties for the \Rt\ probability and \avpt\ distributions as a function of \Rt\ are summarized in Tab.~\ref{tab:sysRT}. The overall systematic uncertainty is calculated by summing the different contributions in quadrature.  
	\begin{table*}[htbp]
	   	\caption{Systematic uncertainties for \Rt probability and \avpt\ distributions as a function of \Rt.}
	\label{tab:sysRT}
	\begin{center}
	\footnotesize
	\begin{tabular}{l|c|c}
	 & \textbf{$R_{\rm T}$ probability distribution} & \textbf{$\langle \pT \rangle$ vs. $R_{\rm T}$ }  \\
	\hline
	\textbf{Unfolding}   & $\pm$ 3.1\%  &  -    \\
	\textbf{Z vertex cut}  & $\pm$   2.4\%  &     $\pm$ 0.2\%   \\
	\textbf{Minimum of $N_{\rm ch}$ to primary vertex}  & $\pm$ 2.7\%  &     $\pm$ 0.2\%      \\
	\textbf{Track cuts}      &  -  & $\pm$ 0.4\%       \\
	\textbf{ITS-TPC track matching}  & -  & $\pm$ 1.3\%    \\
    \textbf{Residual scale}  & $\pm$ 2.2\% & $\pm$ 0.2\%    \\
     \textbf{MC non-closure}  & $\pm$ 0.4\% &  $\pm$ 1.0\%         \\
     \hline
     \hline
     \textbf{Total uncertainty}  & $\pm$ 5.3\% &  $\pm$ 1.6\%         \\
	
	\end{tabular}	
	\end{center}
	\end{table*}



\section{Results and discussion}
\label{sec:results}
In the following the results for the charged-particle number and summed-$\pt$ densities in three azimuthal regions are reported and discussed. Then the results for the \Rt\ probability and \avpt\ vs. \Rt\ distributions in the Transverse region are presented. The experimental results are compared to model calculations using \textsc{Pythia}~8 and \textsc{Epos} LHC. 

\subsection{Charged-particle number density \Nch and $\sum \pt$ distributions}
\begin{figure*}[hbtp]
 \begin{center}
  \includegraphics[width= 0.48\textwidth]{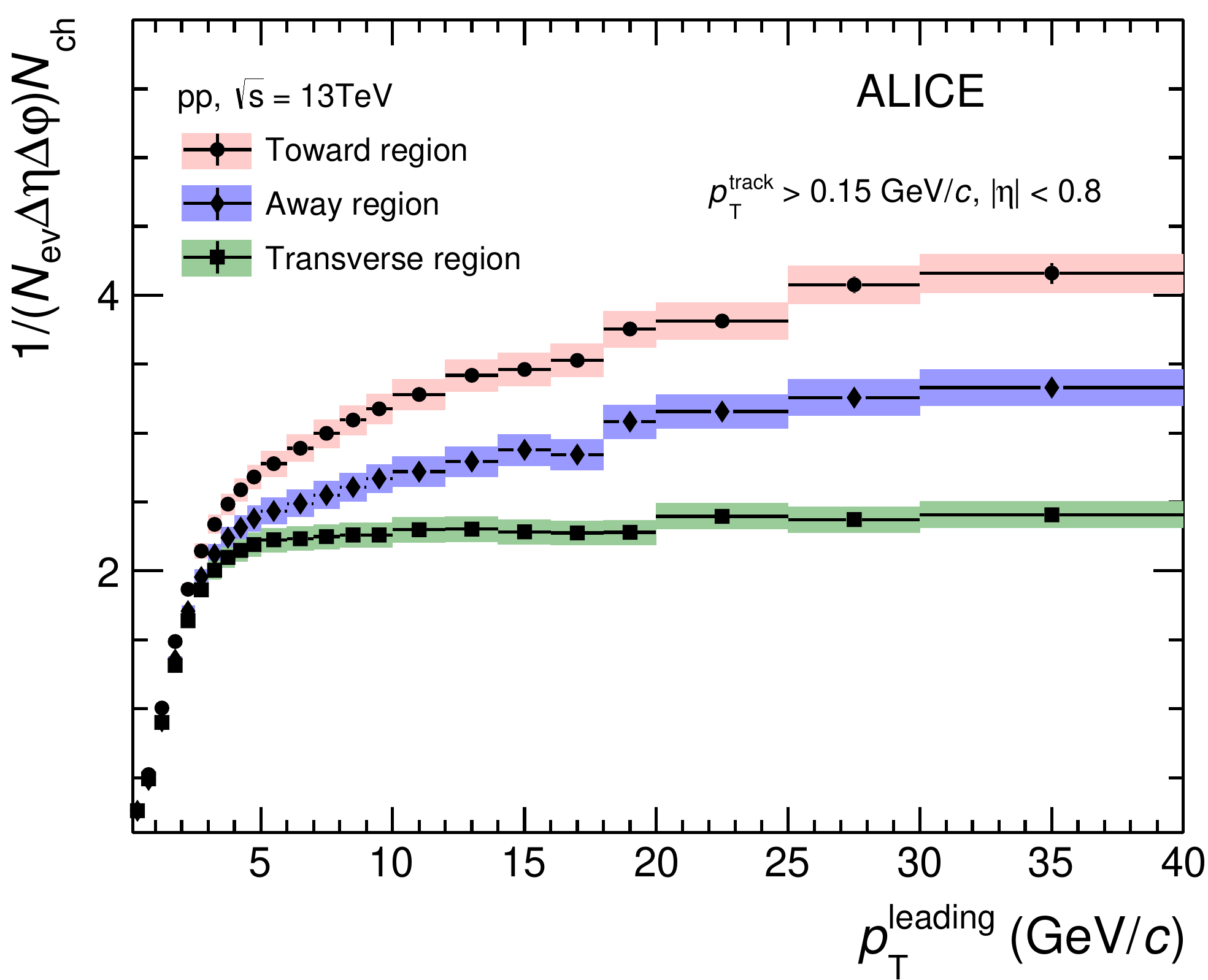}
  \includegraphics[width= 0.48\textwidth]{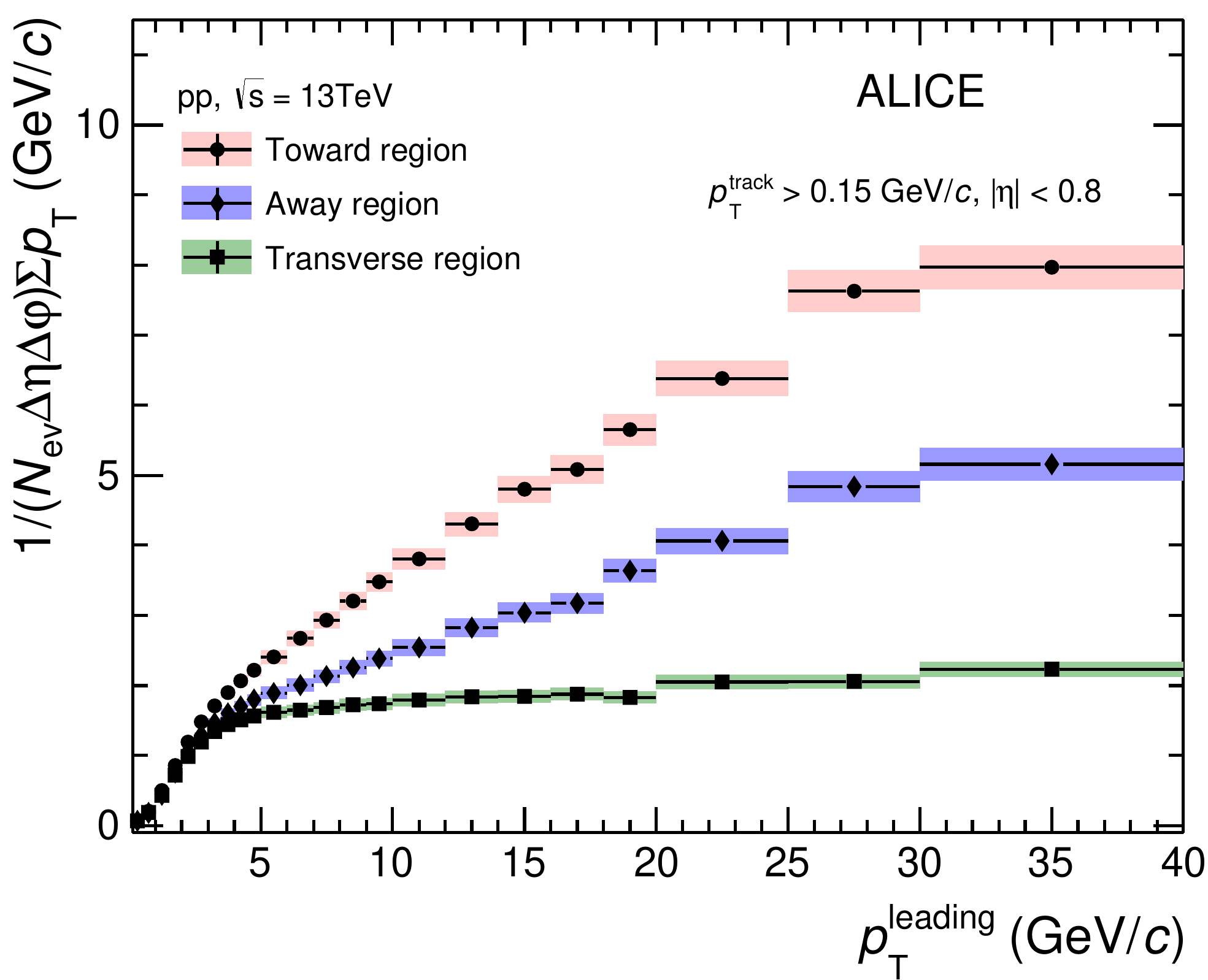}
 \end{center}
 \caption{Number density $\Nch$ (left) and  $\sum \pt$ (right) distributions as a function of \ptlead~in Toward, Transverse, and Away regions for $\ptmin > 0.15 \; \gmom $.  
 The shaded areas represent the systematic uncertainties and vertical error bars indicate statistical uncertainties.
 }
 \label{fig:diffRegion}
\end{figure*}

\begin{figure*}[htbp]
 \begin{center}
 \includegraphics[width=0.45\textwidth]{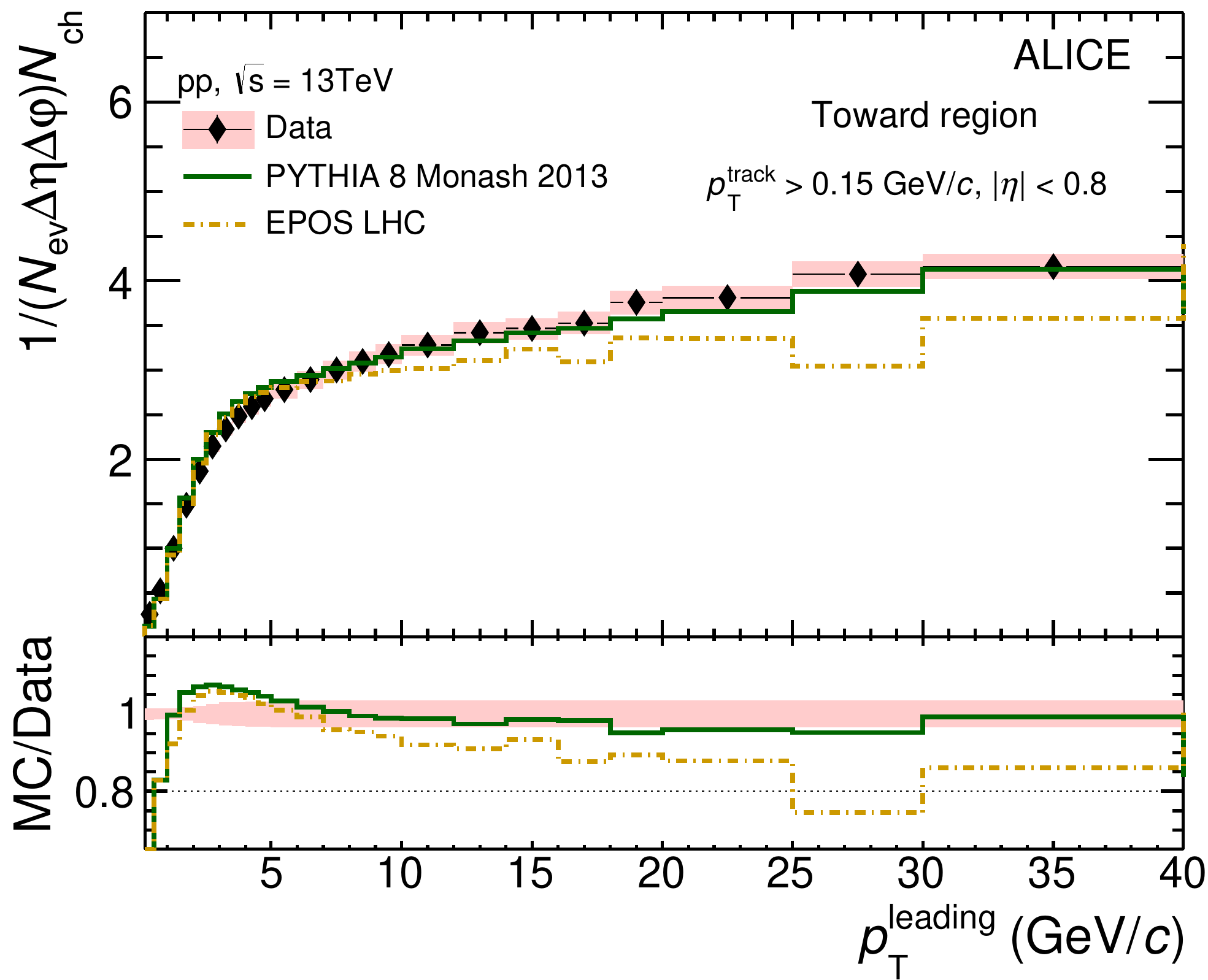}
 \includegraphics[width=0.45\textwidth]{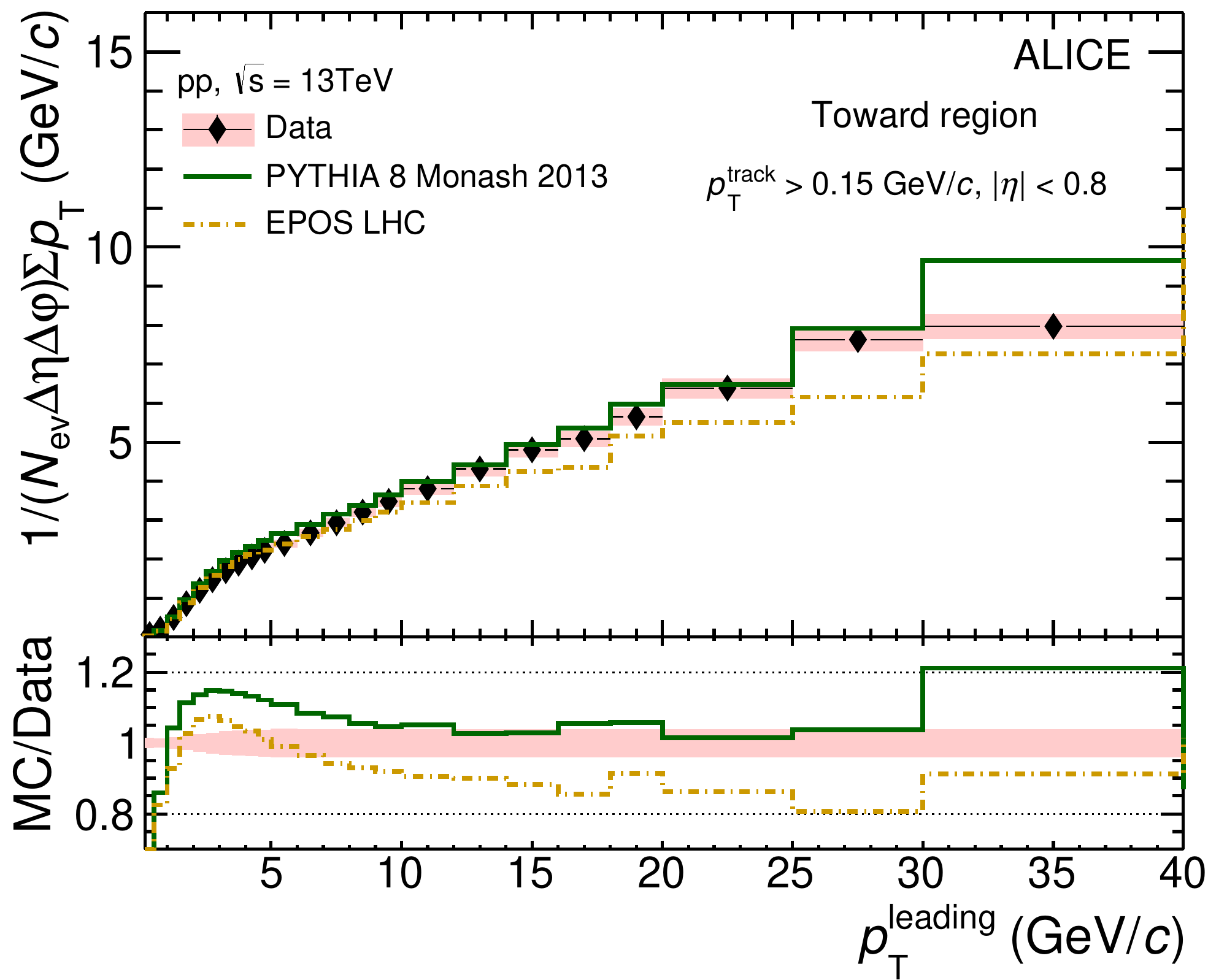}
 \includegraphics[width=0.45\textwidth]{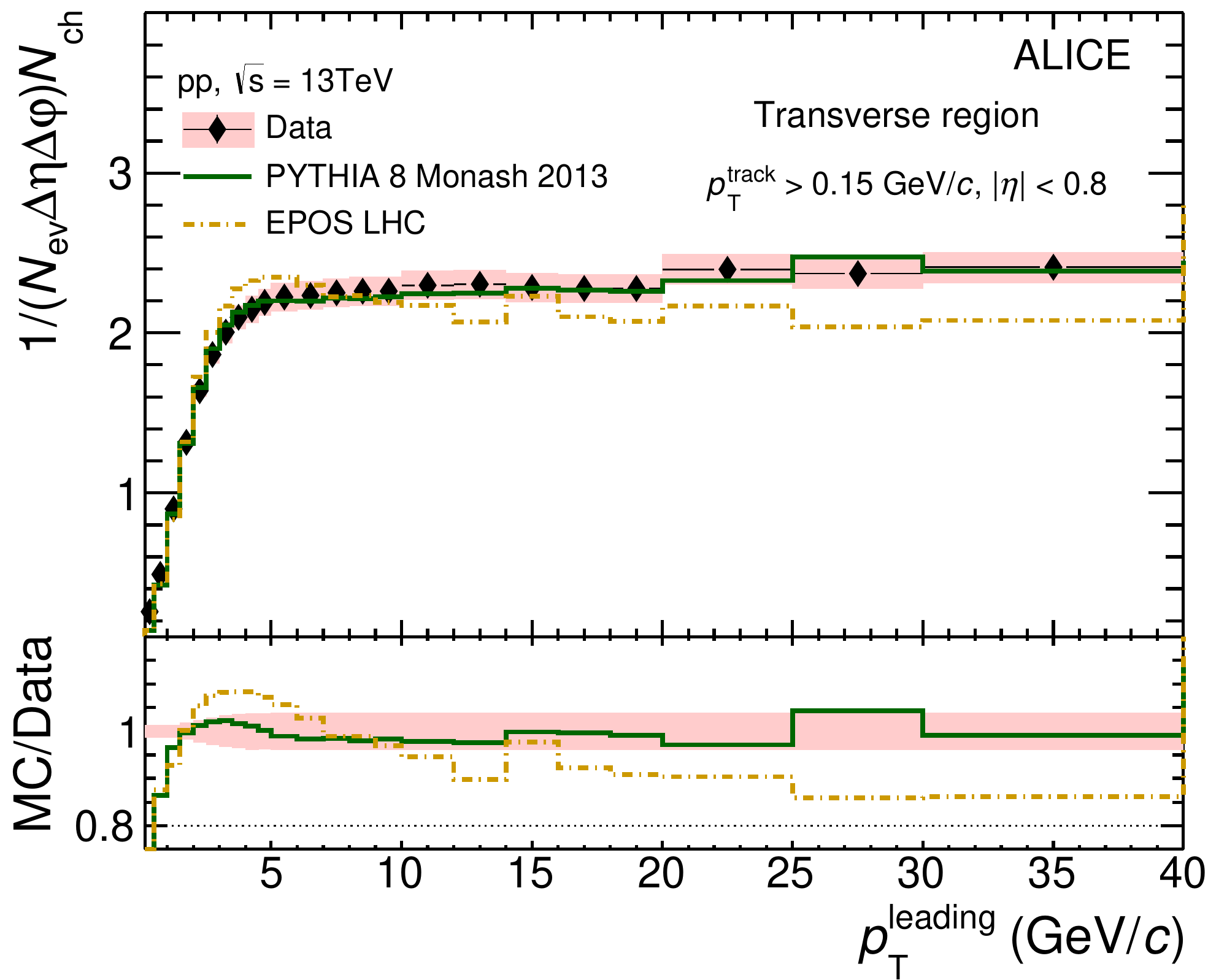}
 \includegraphics[width=0.45\textwidth]{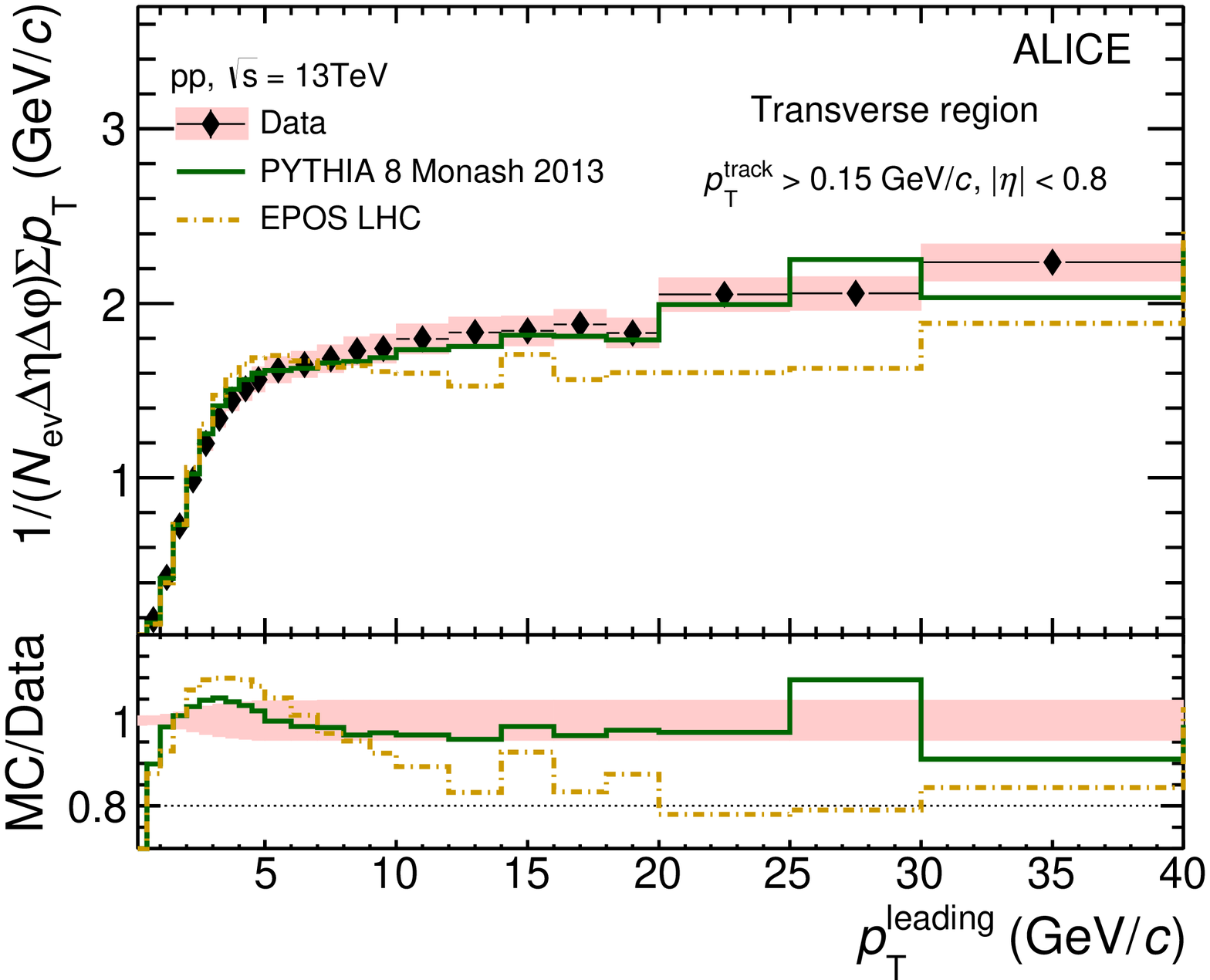}
 \includegraphics[width=0.45\textwidth]{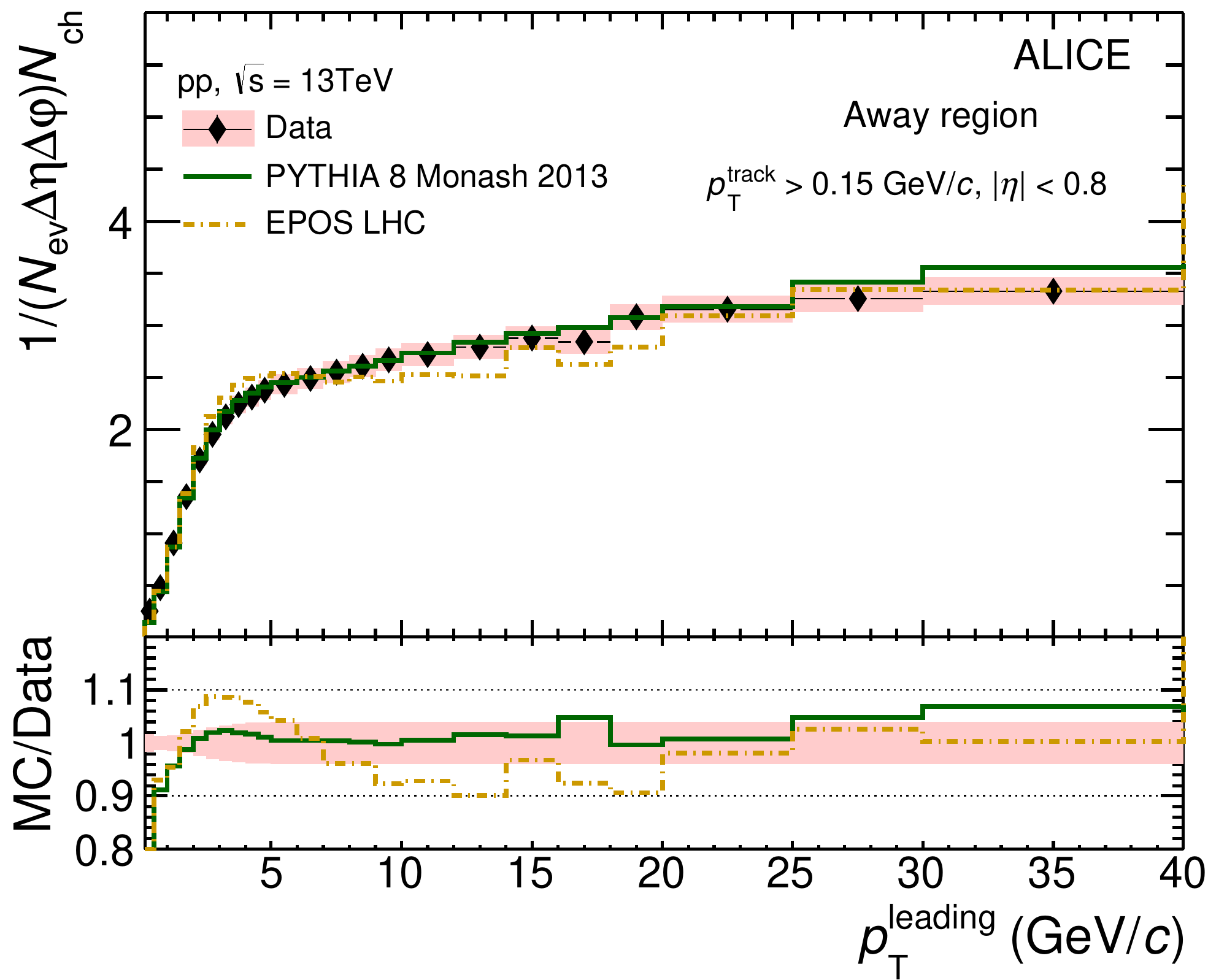}
 \includegraphics[width=0.45\textwidth]{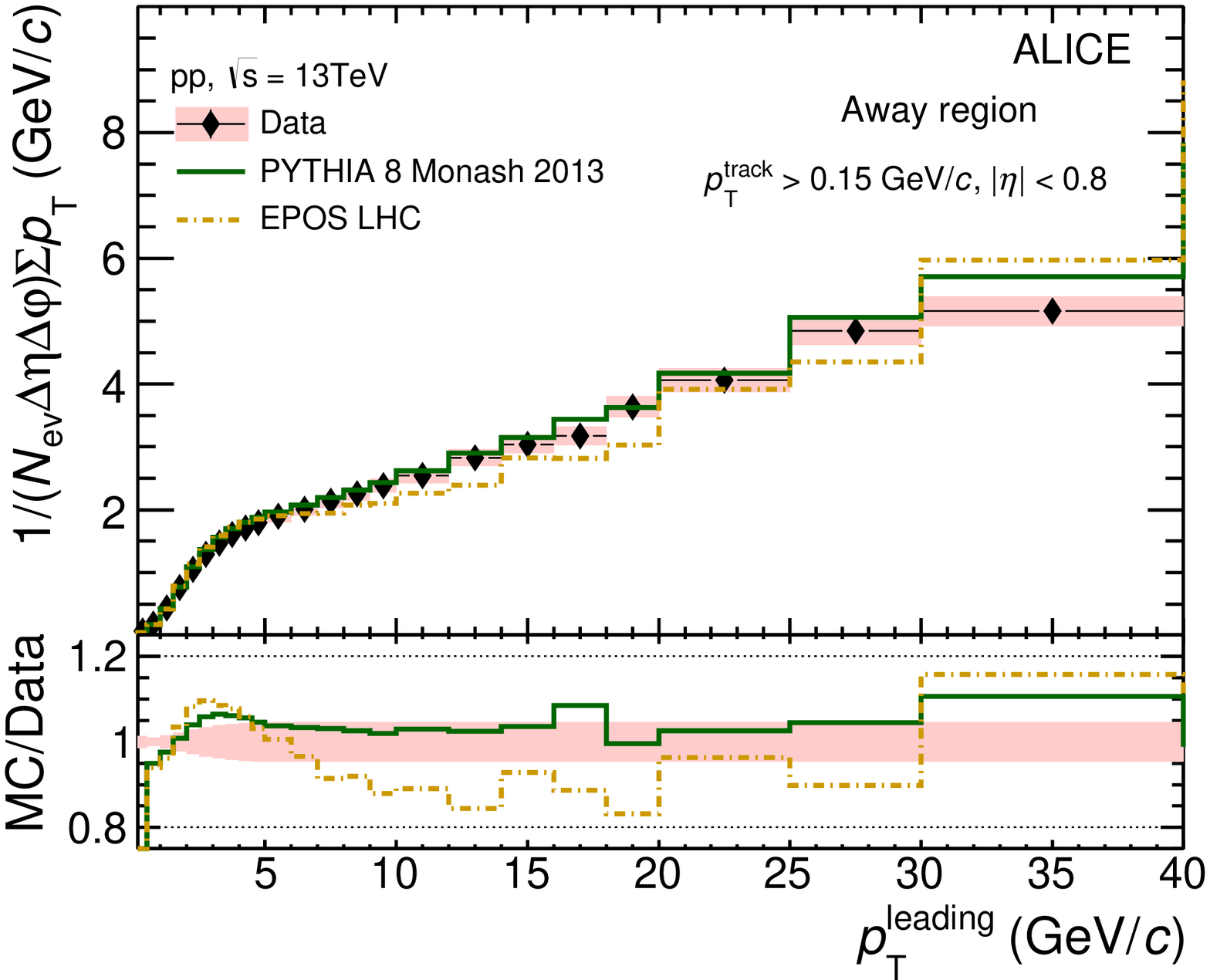}
 \end{center}
 \caption{Number density $\Nch$ (left) and  $\sum \pt$ (right) distributions as a function of \ptlead\, along with the MC simulations in Toward (top), Transverse (middle), and Away (bottom) regions for the threshold of $\ptmin > 0.15 \; \gmom $. The shaded areas in the upper panels represent the systematic uncertainties and vertical error bars indicate statistical uncertainties for the data. In the lower panels, the shaded areas are the sum in quadrature of statistical and systematic uncertainties from upper panels. No uncertainties are shown for the MC simulations.}
 \label{fig: density_015}
\end{figure*}

Figure~\ref{fig:diffRegion} shows the average charged-particle number and summed-$\pt$ densities as a function of $\ptlead$, in the Toward, Transverse, 
and Away regions for the transverse momentum threshold requirement of  $\ptmin\ > 0.15 \; \gmom$. 
The averaged charged-particle number and summed-$\pt$ densities as a function of $\ptlead$ using different $\pt$ thresholds for the associated particles, $\ptmin >$ 0.5 and 1.0 $\gevc$, are presented in Appendix~\ref{app:appendix}.
Figure~\ref{fig: density_015} shows the averaged charged-particle number and summed-$\pt$ densities as a function of $\ptlead$ in the Toward, Transverse, 
and Away regions for the transverse momentum threshold requirement of  $\ptmin\ > 0.15 \; \gmom$, and the comparison to MC results. 
The $\ptlead$ dependence in all regions show a similar behavior. At low $\ptlead$, there is a steep rise in event activity followed by a change to a smaller gradient at $\ptlead \sim 5 \; \gmom$, the plateau region. Above this value in the Transverse region, and in particular for the number density, the event activity becomes almost independent of $\ptlead$, while it continues to rise in the Toward and Away regions. 

In the MPI implementation of \textsc{Pythia}, the average number of hard scatterings per event depends strongly on the impact parameter. Conversely, tagging hard scatterings with 
high-$\pt$ particles biases the events towards lower impact parameter and, hence, higher 
event activity. The change of slope observed in data corresponds to the transverse momentum where the leading particle is dominantly produced by rare hard scatterings, where the average yield per event for such a process is
$\ll 1$.  
This is plausible, in PYTHIA, since requiring particles with lower $\ptlead$ which are produced in almost every parton-parton interaction there cannot be a significant bias on the number of MPI.
The continuous rise observed for the Toward and Away regions can be attributed to particles not only from the UE but also to the contribution of fragments 
from hard scatterings, which are mainly back-to-back in azimuth. 
The contribution from fragments increases with 
\ptlead\, causing the rise of event activity. In contrast, only a small number of 
fragments enter the Transverse region. In addition to the contributions from MPI uncorrelated with 
the hardest scattering, this region contains particles originating from initial-state 
radiation and hard scatterings which produce more than two jets, causing the slow rise of event activity with \ptlead\ observed in the jet plateau range.

\begin{figure*}[hbtp]
 \begin{center}
  \includegraphics[width= 0.47\textwidth]{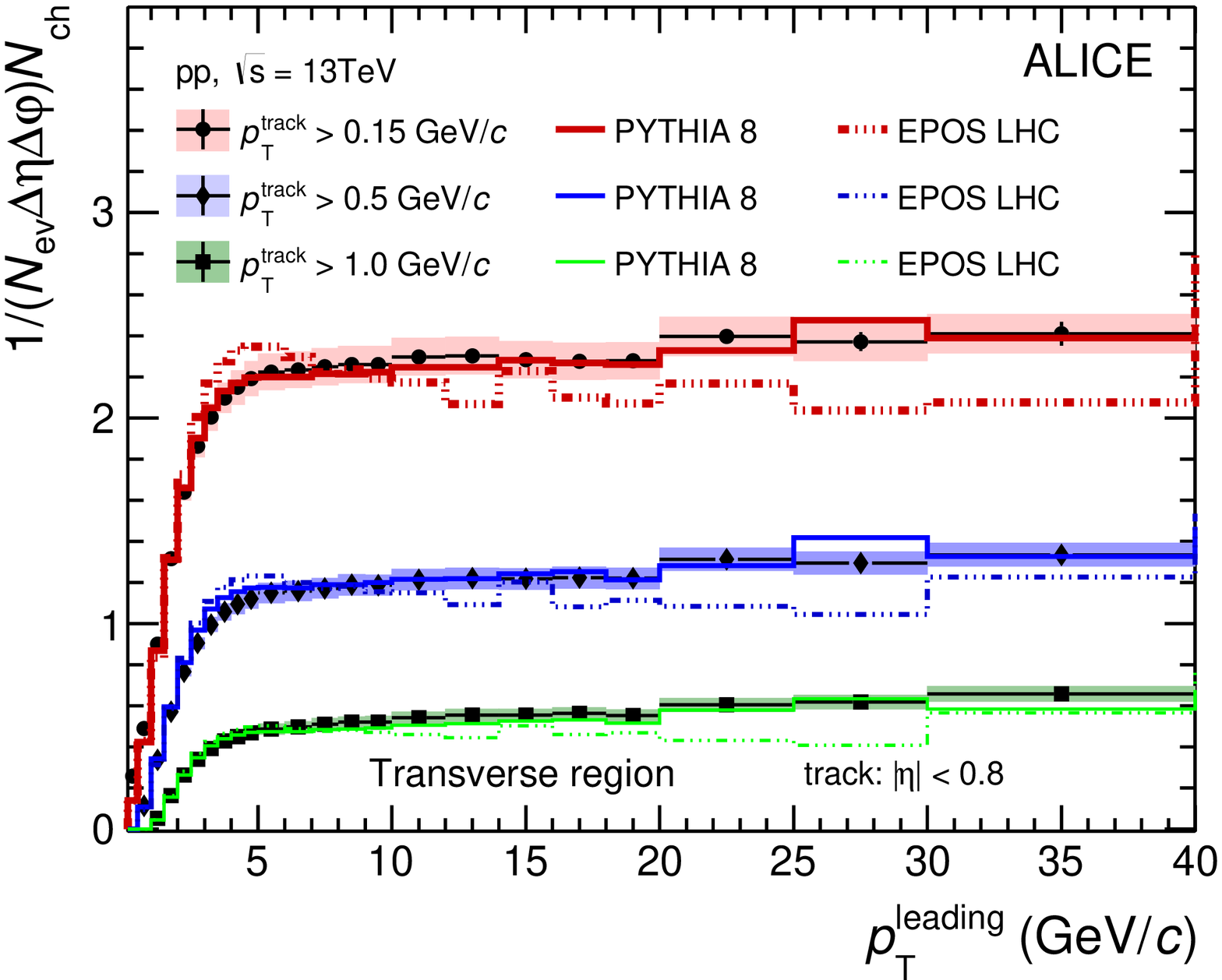}
  \includegraphics[width= 0.5\textwidth]{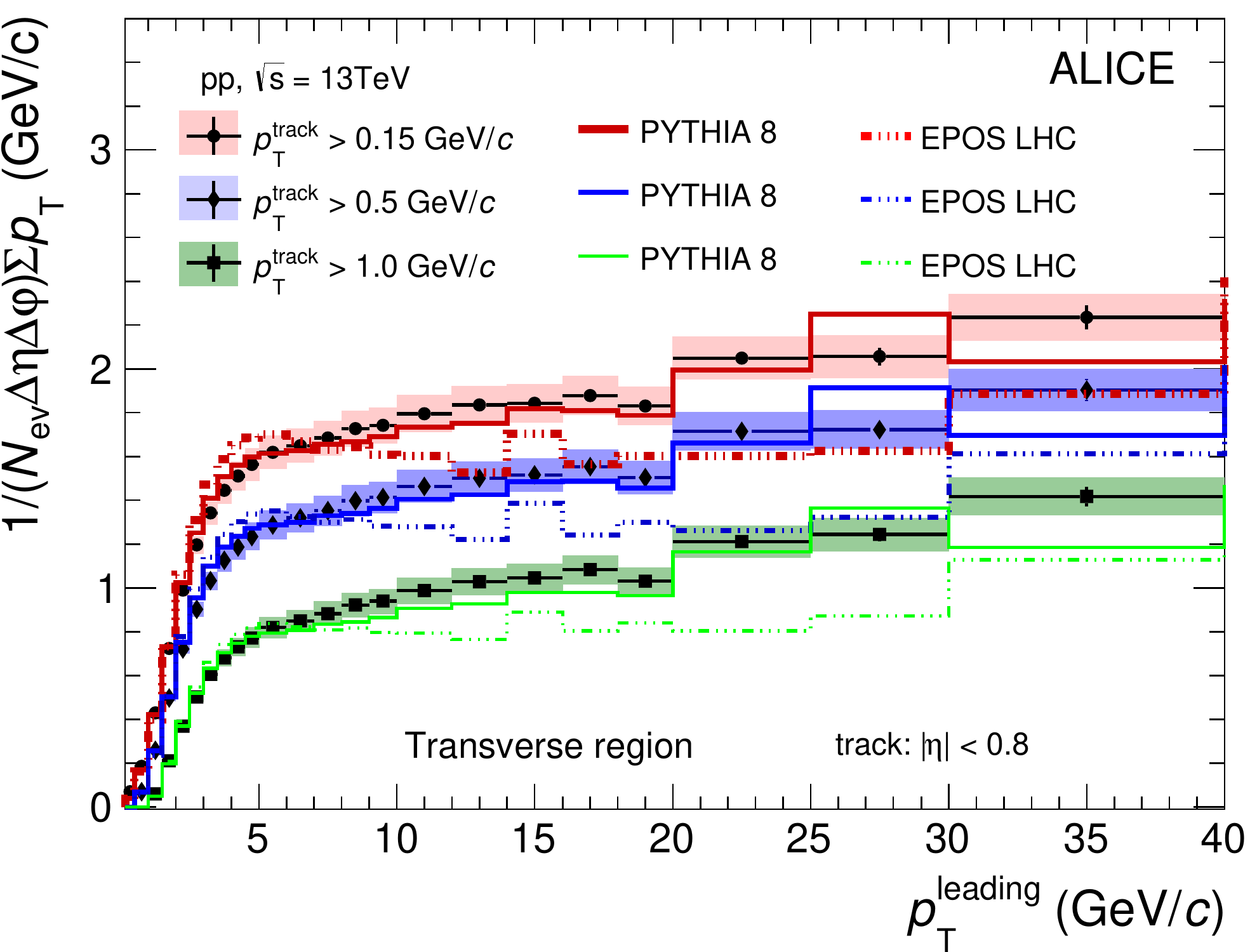}
 \end{center}
 \caption{Number density $\Nch$ (left) and  $\sum \pt$ (right) distributions as a function of \ptlead\, along with the MC simulations in Transverse region for three transverse momentum thresholds of $\ptmin\ > 0.15$, $0.5$, and $1.0 \; \gmom$.
 The shaded areas represent the systematic uncertainties and vertical error bars indicate statistical uncertainties for the data. No uncertainties are shown for the MC simulations.
 }
 \label{fig:diffPtcut}
\end{figure*}

Figure~\ref{fig:diffPtcut}  compares the number density and $\sum \pt$ distributions as a function of \ptlead\  
in the Transverse region for the three threshold selections $\ptmin\ > 0.15$, $0.5$, and $1.0 \; \gmom$. 
In the plateau region, increasing the cut from $\ptmin > 0.15 \; \gmom $ to $\ptmin > 1.0 \; \gmom $ reduces the number density by almost a factor of 4. The relative slope of the distributions in the pedestal region slightly increases with the $\ptmin$ threshold, indicating an increased contribution of correlated hard processes (initial-state radiation) to the Transverse region. This shows that the highest sensitivity to the UE is obtained using the lowest $\pT$ threshold.

Figures~\ref{fig: density_015} and~\ref{fig:diffPtcut}
also compare the number density $\Nch$ and  $\sum \pt$ densities to the 
results of \textsc{Pythia}~8 and \textsc{Epos} LHC calculations.
\textsc{Pythia}~8 describes the plateau in the Transverse region quite well, while \textsc{Epos} LHC underpredicts the densities in this region (as well as in the Toward region) by about 20\%. When increasing the $\ptmin$ cut, the agreement between data and \textsc{Pythia}~8 in the Toward and Away regions becomes slightly worse (see Appendix~\ref{app:appendix}), with the largest discrepancies appearing at low \ptlead~in the Toward region. In general \textsc{Epos} LHC fails to reproduce the experimental data in most regions. This can be attributed to the underprediction of the number of hard scatterings in the model.  The issue is expected to be solved in EPOS 3 using a new variable saturation scale~\cite{Werner:2013vba, Pierog:2019opp}.



\begin{figure*}[hbtp]
 \begin{center}
  \includegraphics[width= 0.49\textwidth]{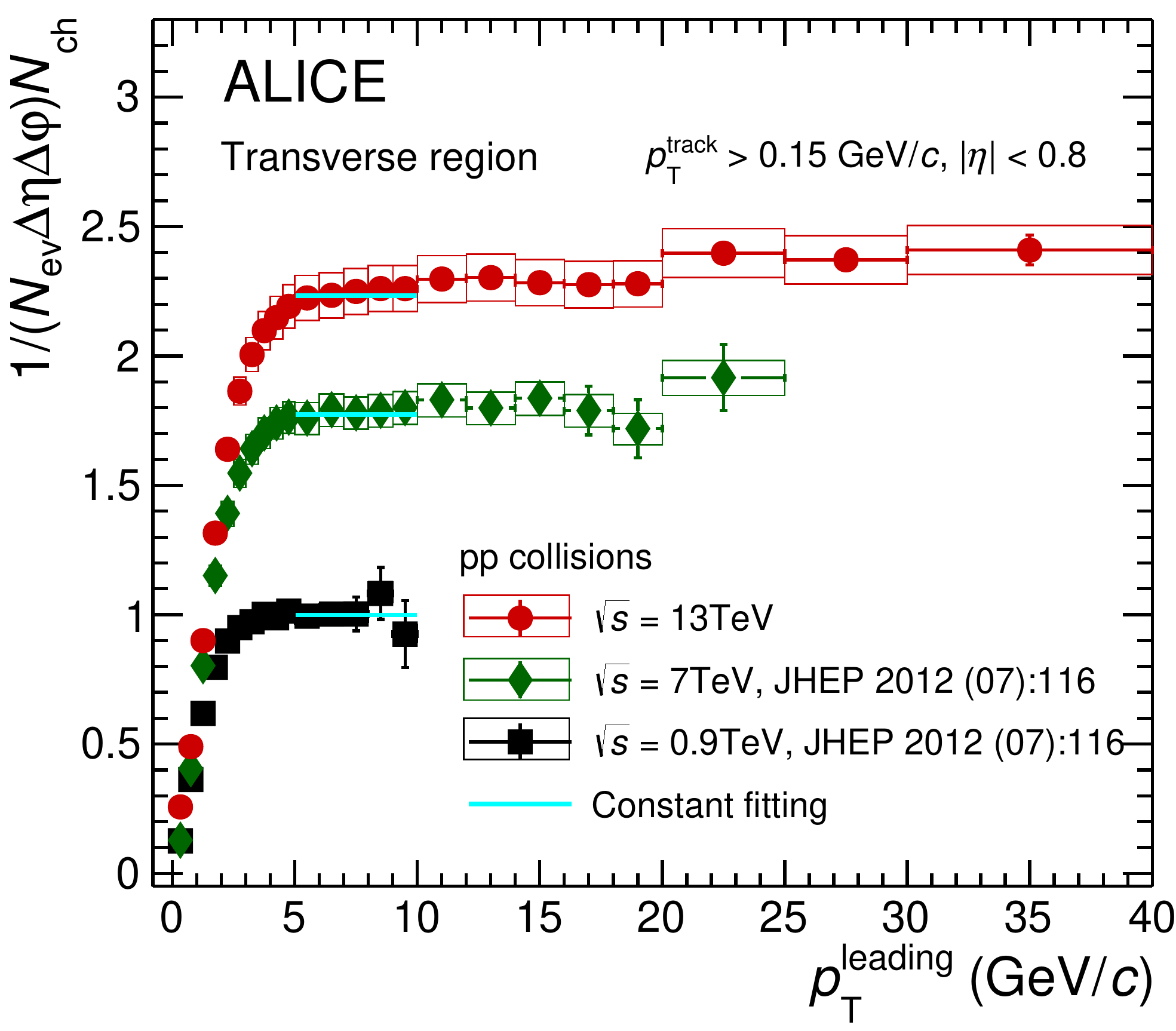}
  \includegraphics[width= 0.49\textwidth]{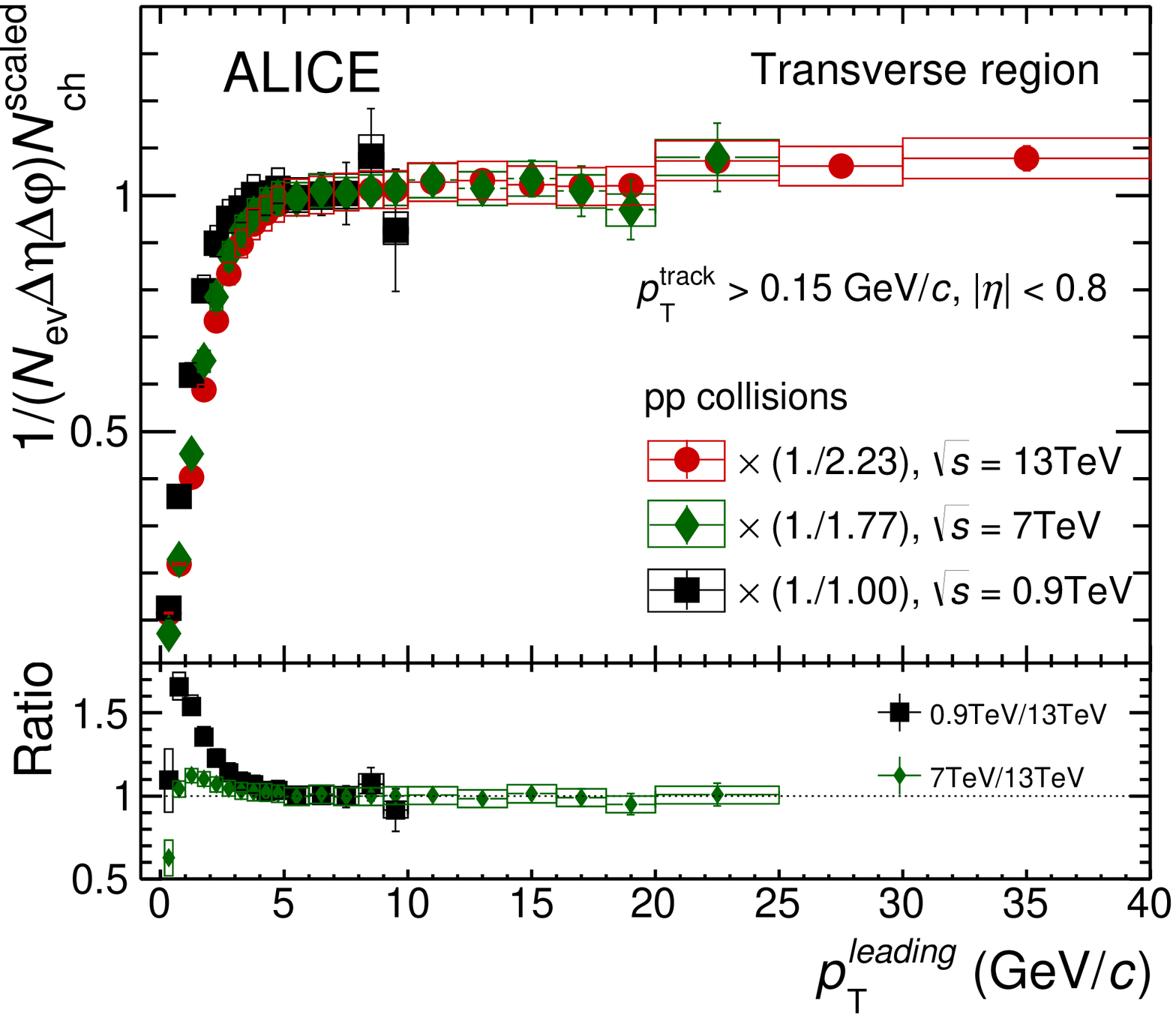}
 \end{center}
 \caption{Left: Number density $\Nch$ in the Transverse region as a function of \ptlead\ 
 ($\ptmin > 0.15 \; \gmom$ threshold) for $\sqrt{s} = $ 0.9, 7 and 13~$\TeV$. A constant function is used to fit the data in the range $5 < \ptlead < ~ 10 \; \gmom$ and the results are shown as solid lines.
 Right: Number densities $\Nch$ scaled by the pedestal values obtained from the fit in order to compare the shapes. The open boxes represent the systematic uncertainties and vertical error bars indicate statistical uncertainties.
 }
 \label{fig: diffScom}
\end{figure*}

Figure~\ref{fig: diffScom} (left) shows the comparison of the results obtained at $\sqrt{s} = 13 \; \TeV$ to the ones obtained at lower collision energies, $\sqrt{s} = 0.9 \ $ and $7 \; \TeV$~\cite{A51_ALICE_UE}, in the Transverse region. Between the two higher energies, the number density in the plateau increases by about $30\%$.
%
More information about the $\sqrt{s}$-dependence in the Transverse region can be obtained by comparing the shapes of the number density vs. $\ptlead$.
To this end, the height of the plateau for different collision energies is quantified by fitting a constant function in the range $5 < \ptlead\ < 10 \; \gmom$, shown as lines in Fig.~\ref{fig: diffScom} (left). 
The fitting range was restricted to the common range in order to be consistent with the procedure used for the measurements at lower $\sqrt{s}$. Larger fitting ranges were also considered and consistent results were obtained. 
The shapes of the particle densities as a function of \ptlead\ are then compared after dividing the densities by the height of the plateau. The results are shown in Fig.~\ref{fig: diffScom} (right).  For the two higher energies the coverage extends beyond the fitting range, i.e. to $\pT > 10 \; \gmom $. In this range the densities agree within the statistical and systematic uncertainties.
In the region of the rise ($\pt < 5 \; \gmom$) one observes a clear ordering among the three collision energies, the lowest energy having the highest density relative to the plateau. 
At lower $\sqrts $ the plateau starts at a slightly lower $\ptlead $.

%
    
\begin{figure*}[hbtp]
 \begin{center}
  \includegraphics[width= 0.65\textwidth]{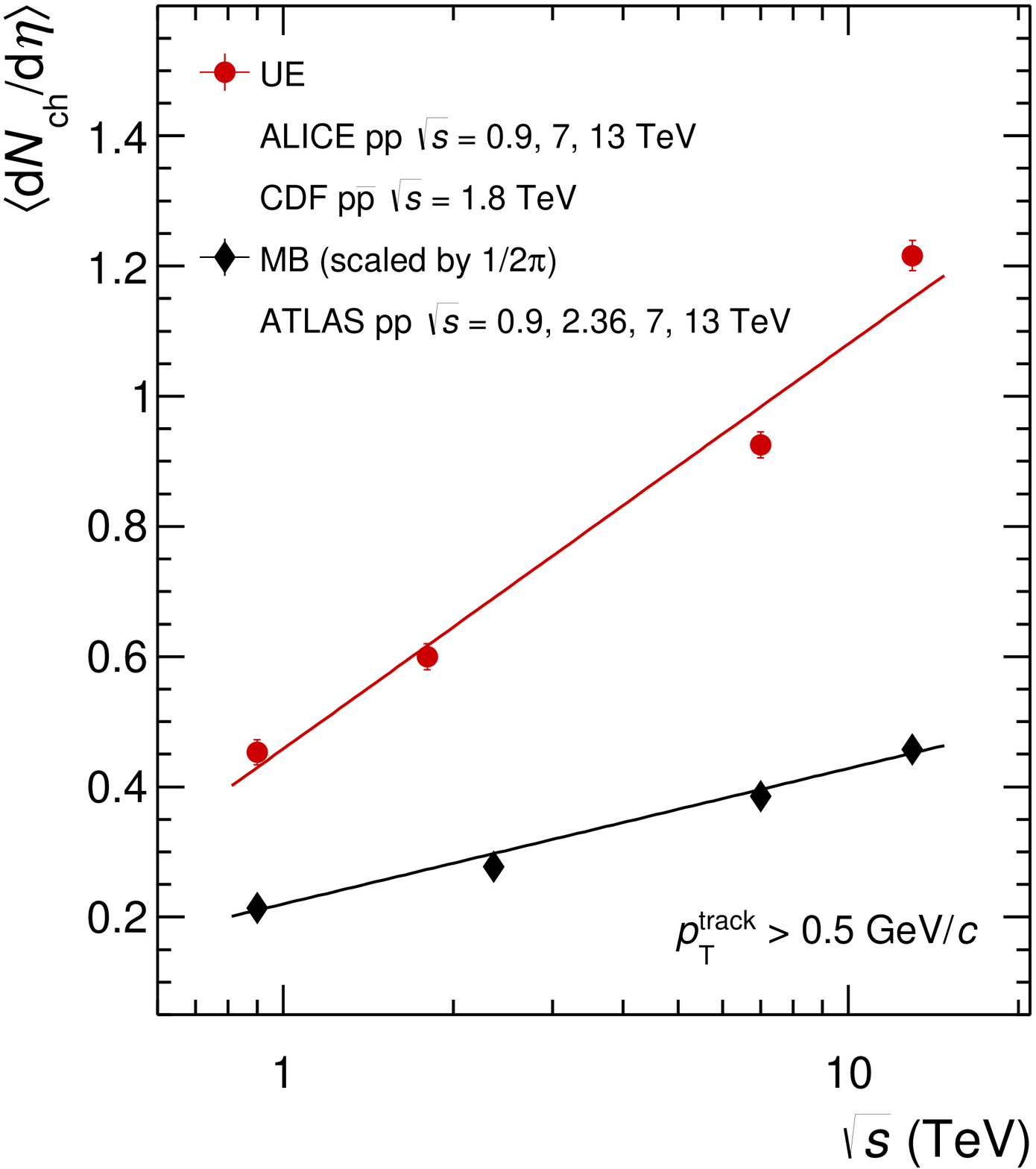}
 \end{center}
 \caption{Comparison of number density in the plateau of the Transverse region with lower energy data taken from~\cite{cdf1, A51_ALICE_UE} and $dN_{\rm ch}/d\eta$ in minimum-bias events (scaled by $1/2\pi$)~\cite{atlasdndeta}. Both are for charged-particles with $\pt  >$ 0.5~\gmom. Error bars represent statistical and systematic uncertainties summed in quadrature. The straight lines show the results of fitting data points with the functional form $a+b\ln{s}$. }
 \label{fig: ueCollS}
\end{figure*}

Figure~\ref{fig: ueCollS} shows the $\sqrts$-dependence of the number density of the plateau in the Transverse region for $\ptmin > 0.5~\gmom$, from a fitting of a constant function in the \ptlead~range $5 < \ptlead < ~ 10 \; \gmom$. The lower energy data are taken from ALICE~~\cite{A51_ALICE_UE} and CDF~\cite{cdf1} measurements. 
It is compared with the midrapidity charged-particle density ${\rm d}N_{\rm ch}/{\rm d}\eta|_{\eta=0}$ of charged-particles with $\pt >$~0.5~\gmom\ in MB events also requiring at least one charged particle in $|\eta| <$ 2.5 (scaled by $1/2\pi$)~\cite{atlasdndeta}.
The UE activity in the plateau region is more than a factor of two higher than ${\rm d}N_{\rm ch}/{\rm d}\eta$.
Both are consistent with a logarithmic dependence on collision energy.
Between $\sqrts =$ 0.9 and 13~$\TeV$ ${\rm d}N_{\rm ch}/{\rm d}\eta$ increases by approximately a 
factor of 2.1 whereas the increase of the UE activity is $30\%$ larger,
confirming the trend observed previously for collision energies up to $7 \; \TeV$.

\begin{figure*}[hbtp]
 \begin{center}
  \includegraphics[width= 0.46\textwidth]{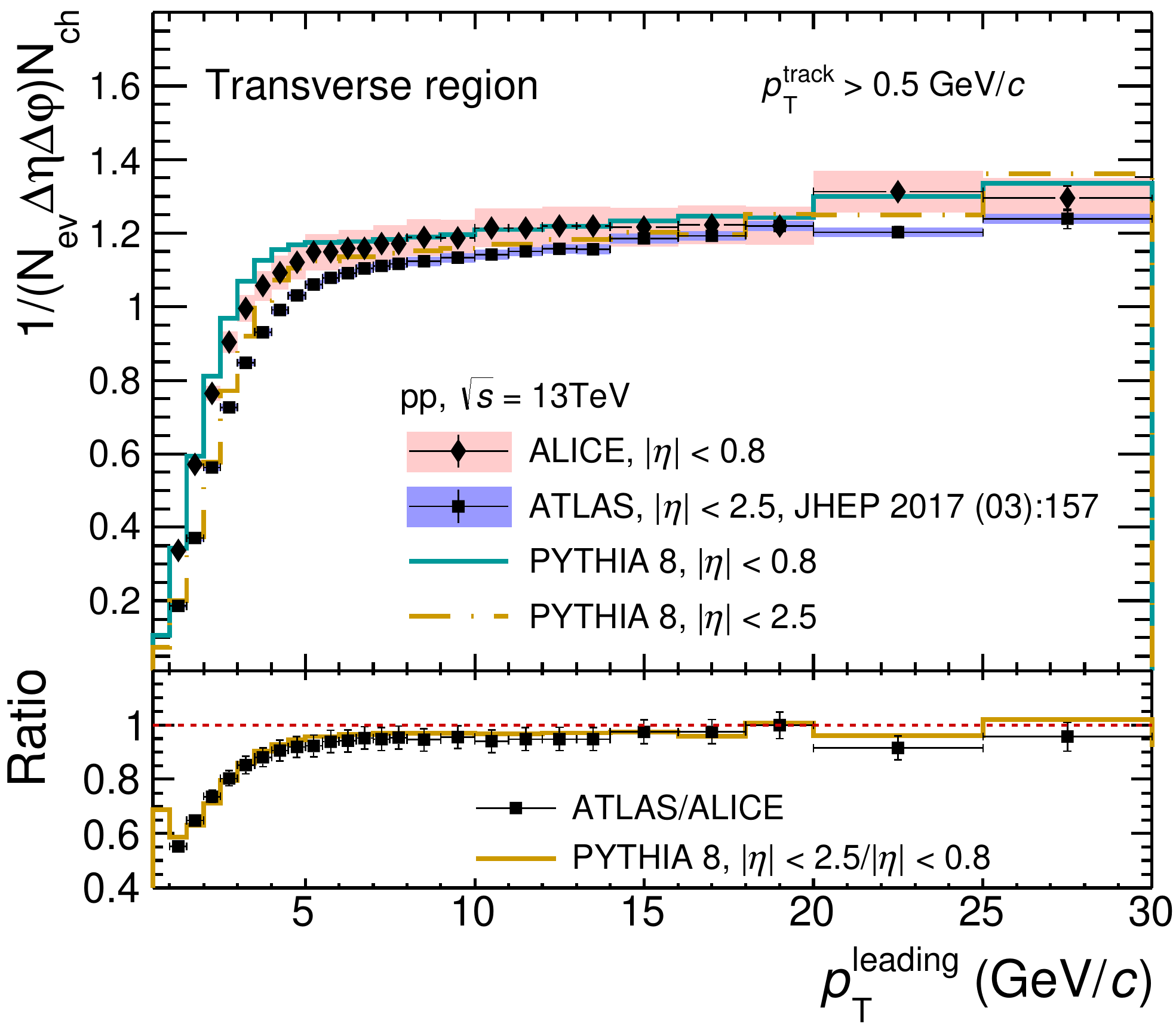}
  \includegraphics[width= 0.46\textwidth]{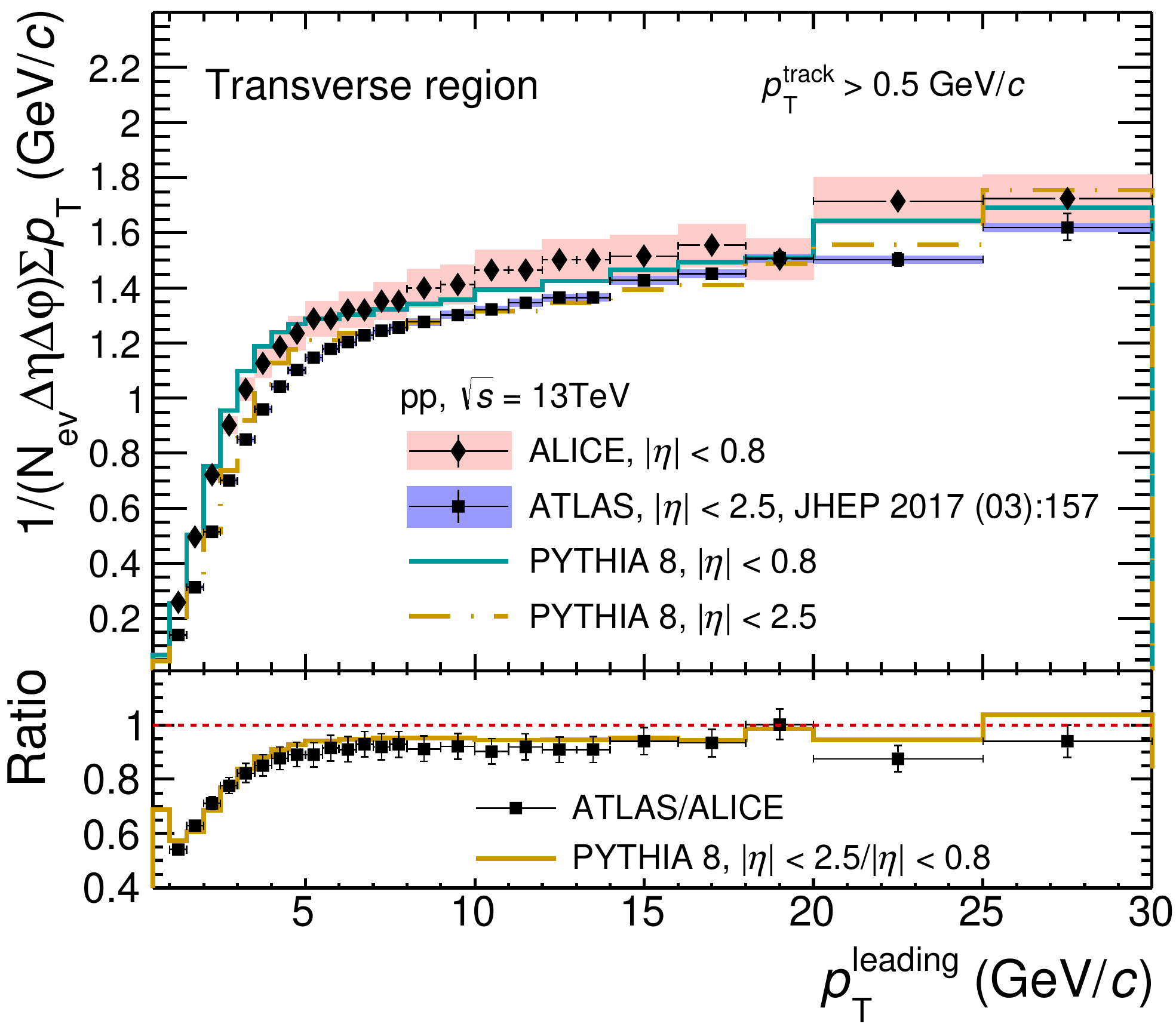}
 \end{center}
 \caption{Number density $\Nch$ (left) and $\sum \pt$  density (right) in the Transverse region for $\ptmin > 0.5$ ~\gmom\  at $\sqrt{s}=13~\TeV$ from ALICE ($|\eta| < 0.8$) and ATLAS ($|\eta| < 2.5$) measurements~\cite{JHEP03_2017_157}. The results are compared to \textsc{Pythia}~8 Monash-2013 calculations. The shaded areas represent the systematic uncertainties and vertical error bars indicate statistical uncertainties for the data. No uncertainties are shown for the MC simulations.}
 \label{fig: AtlasComp}
\end{figure*}

ATLAS has published similar UE results measured in a wider rapidity acceptance of $|\eta| < 2.5$ and using a threshold of $\ptmin > 0.5 \; \gmom$~\cite{JHEP03_2017_157}. 
Since jets have a finite extension in $\eta$-$\varphi$ space,
the larger acceptance allows more particles from the leading jet fragmentation and in particular, from the away-side partner jet subject to a pseudorapidity swing, to enter the measurement and, hence,  the results in the Toward and Away regions are not directly comparable between the two experiments. 
Notably, the smaller acceptance obscures an interesting feature observed by ATLAS: for $\ptlead > 7$~\gmom\ the Away region has a higher charged-particle multiplicity density than the Toward region, despite not containing the highest-$\pt$ charged particle. In the ALICE measurement the Toward region has always the higher multiplicity density (see Fig. \ref{fig:diffRegion}). 
However, when comparing the distributions from the Transverse region one observes good agreement (Fig.~\ref{fig: AtlasComp}) in the plateau region. This indicates that the UE activity does not depend strongly on the rapidity coverage and that the fact that in some
cases particles with $\pt > \ptlead$ are outside the acceptance does not have a strong effect on the measurement.
For the lower acceptance used in ALICE, the plateau starts at a slightly lower $\ptlead$. As a consequence, in the region of the steep rise for $\ptlead < 5 \; \gmom$, the ratio between the densities for the higher and lower $\eta$ acceptance increases strongly with $\ptlead$. 
In MPI-based models, like the one implemented in \textsc{Pythia}~8, the onset of the plateau is reached when the per-event probability to find a leading particle of a given $\ptlead$ is much less than unity. Decreasing the acceptance or, as discussed above, lowering the collision energy would move the onset of the plateau to smaller $\ptlead$, which is in  agreement with our observations (see \textsc{Pythia}~8 comparison in Fig.~\ref{fig: AtlasComp}).

\subsection{Relative Transverse activity classifier \Rt distributions}
The  \Rt analysis is performed using a track transverse momentum threshold of $\ptmin\ > 0.15 \; \gmom$ and by selecting events in the plateau region  ($\ptlead > 5 $~GeV/$c$).
Here the $R_{\rm T}$ probability distribution and the mean charged-particle $\pT$ as a function of $\Rt$, in the Transverse region, are reported.



%
	
\begin{figure*}[hbtp]
 \begin{center}
  \includegraphics[width= 0.65\textwidth]{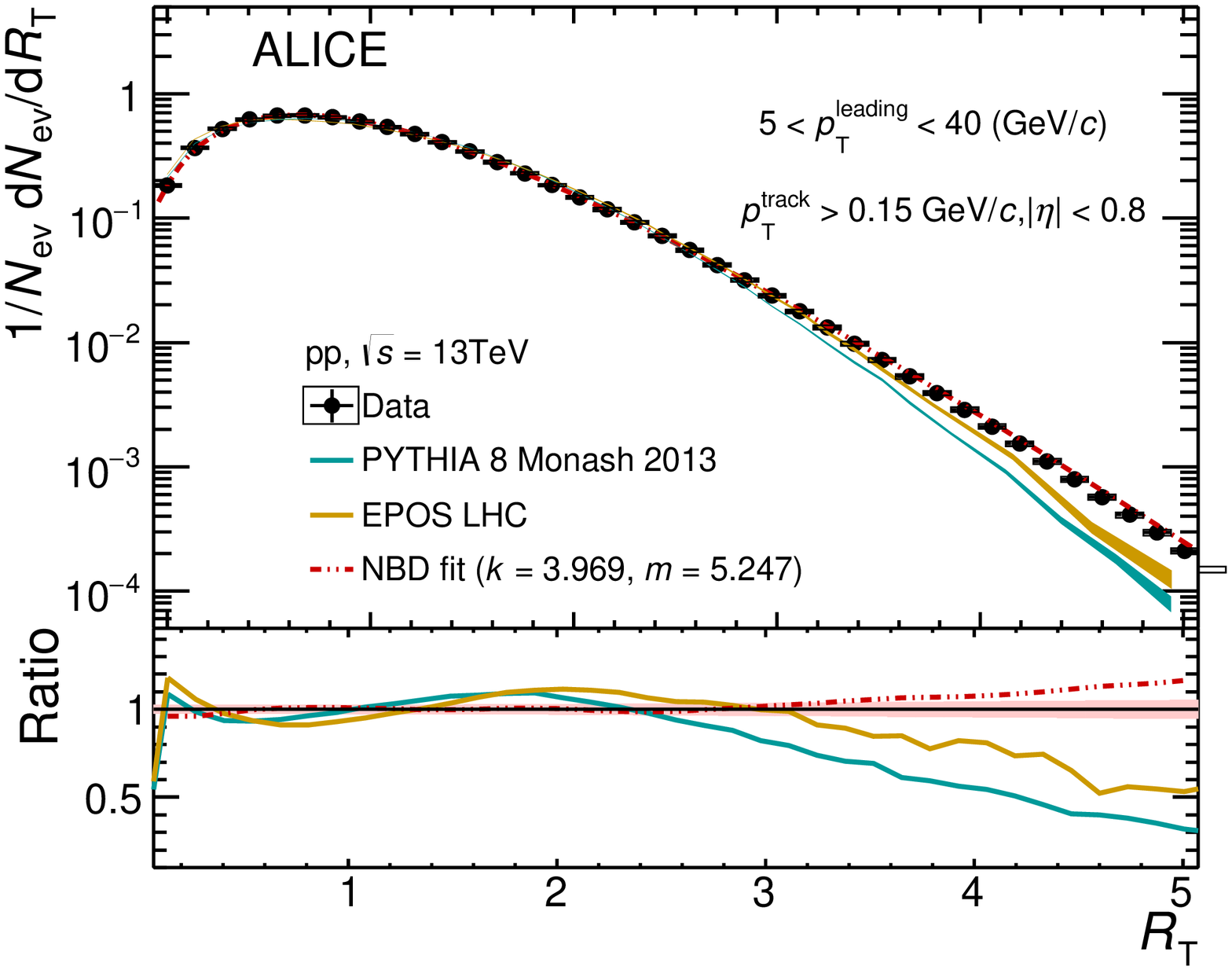}
 \end{center}
 \caption{
 \Rt probability distribution in the Transverse region for $\ptmin > 0.15 \; \gmom $ and $|\eta | < 0.8$. 
 The result (solid circles) is compared to the \textsc{Pythia}~8 and \textsc{Epos} LHC calculations (lines).
 The red line represents the result of the NBD fit, where the multiplicity is scaled by its mean value, $m$. 
 The parameter $k$ is related to the standard deviation of the distribution via 
 $ \sigma = \sqrt{\frac{1}{m}+\frac{1}{k}} $.
 The open boxes represent the systematic uncertainties and vertical error bars indicate statistical uncertainties for the data. The bands indicate the statistical uncertainties of the MC simulations. The bottom panel shows the ratio between the NBD fit, as well as those of the MC to the data.
 }
 \label{fig:RTDis}
\end{figure*}

The \Rt\ probability distribution is shown in Fig.~\ref{fig:RTDis}. The distribution has been fitted by a modified NBD with the multiplicity scaled by its average value, as was done for the measured \Rt observable.
Within the experimental uncertainties, the NBD fit gives a good description of the data up to \Rt = 3, and it slightly overestimates the data with increasing \Rt, by about 14\% at \Rt = 5. 
The distribution is also compared with the calculations from \textsc{Pythia}~8 and \textsc{Epos} LHC. While both models describe the data well in the \Rt\ regions close to the peak at $ \Rt \sim 0.7$, both \textsc{Pythia}~8 and \textsc{Epos} LHC calculations diverge strongly at higher \Rt, and underpredict the \Rt distribution by more than a factor of two for \Rt $>$ 4.
This opens possibilities to study the interplay of components of pp collisions. Detailed MC event generators studies are needed to interpret the mechanisms
responsible for the disagreement at high \Rt values.

\begin{figure*}[hbtp]
 \begin{center}
  \includegraphics[width= 0.65\textwidth]{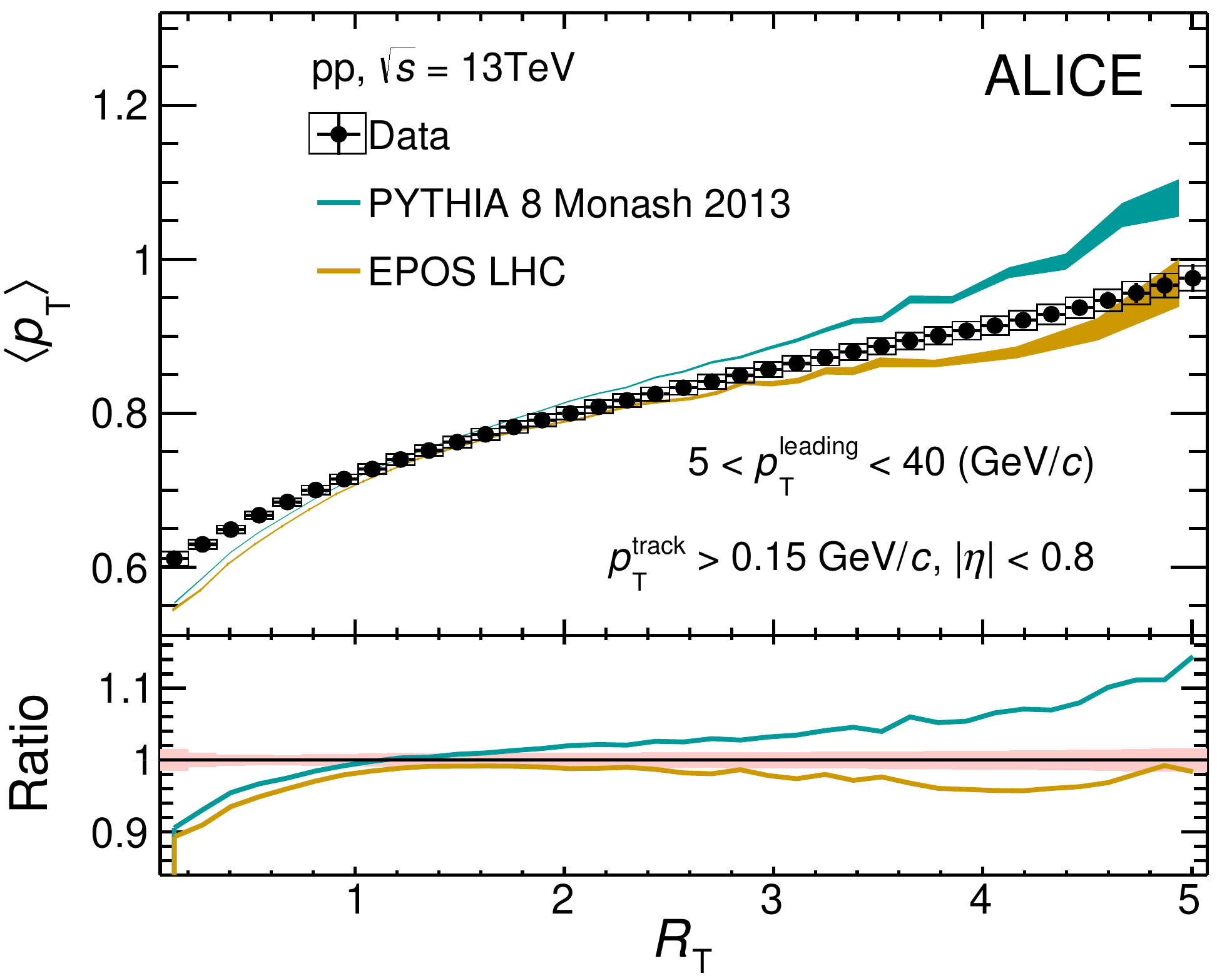}
 \end{center}
 \caption{ $\avpt$ in the Transverse region as a function of \Rt\ for $\ptmin > 0.15 \; \gmom $ and $|\eta | < 0.8$. Data (solid circles) are compared to the results of \textsc{Pythia}~8 and \textsc{Epos} LHC calculations (lines). 
 The open boxes represent the systematic uncertainties and vertical error bars indicate statistical uncertainties for the data. The bands indicate the statistical uncertainties of the MC simulations. The bottom panel shows the ratio of the MC to data.}
 \label{fig:MeanPTRT}
\end{figure*}

The charged-particle $\avpt$ distribution as a function of \Rt\ is shown in
Fig.~\ref{fig:MeanPTRT}. The average transverse momentum rises steadily from $\sim 0.6 \; \gmom$ at low UE multiplicity to $\sim 1 \; \gmom$ for 5 times the mean multiplicity. 
The results are also compared with the \textsc{Pythia}~8 and \textsc{Epos} LHC calculations. 
While the shapes are similar, both models deviate from the measurement by up to 10\%,
in particular at the extremes of the $\Rt$ interval covered by the measurement.
Interestingly, at high multiplicity the deviations have opposite signs for the two models, with \textsc{Pythia}~8 predicting slightly harder and \textsc{Epos} LHC softer transverse activity than seen in data. 

\section{Conclusions}
\label{sec:conclusion}
The UE properties in pp collisions at ${\sqrt{s}=13} ~{\tev}$ have been characterized by measuring the number density $\Nch$ and $\sum \pt$ density distributions of charged particles in three azimuthal regions with respect to the leading charged-particle direction: Toward, Transverse, and Away.
The measurement is performed using charged particles, which have been corrected to the level of primary charged particles. 
The results are compared to previous ALICE measurements in pp collisions at $\sqrts = 0.9$ and $7~\, \tev$. In this work, the kinematic range of the leading particle, $\ptlead$, is extended, and the uncertainties are reduced. An increase of approximately 30\% of the jet pedestal is observed when the pp collision energy increases from $\sqrts = 7$ to $13$ TeV. The UE activity, quantified by the charged-particle density in the jet pedestal range in the Transverse region, shows a stronger increase with $\sqrts$ than the inclusive midrapidity charged-particle density in MB events.  
This is in qualitative agreement with an
increased relative contribution of hard processes to the UE with increasing $\sqrts$.

The Transverse region has been further characterized by the relative transverse UE activity classifier $\Rt$.
Measuring UE quantities versus \Rt\ yields sensitivity to rare events with exceptionally large or small transverse activity with respect to the event-averaged mean. 
The models considered in the paper, \textsc{Pythia}~8 and \textsc{Epos}~LHC, cannot describe the $\Rt$ distribution in the full range covered by the measurements  ($ 0 < \Rt < 5$). 
Moreover, whereas the overall agreement with  \avpt\ measured in the transverse region as a function of \Rt\ is within 10\%, \textsc{Pythia}~8 and \textsc{Epos}~LHC show significant deviations at very low and high $\Rt$.
Compared to data and to each other, these models show a significantly different behavior at high $\Rt$.
This might be a consequence of how each model treats the high-density events. Therefore, the measurements presented here provide new constraints on the models, particularly to their description of MPI.


\newenvironment{acknowledgement}{\relax}{\relax}
\begin{acknowledgement}
\section*{Acknowledgements}

The ALICE Collaboration would like to thank all its engineers and technicians for their invaluable contributions to the construction of the experiment and the CERN accelerator teams for the outstanding performance of the LHC complex.
The ALICE Collaboration gratefully acknowledges the resources and support provided by all Grid centres and the Worldwide LHC Computing Grid (WLCG) collaboration.
The ALICE Collaboration acknowledges the following funding agencies for their support in building and running the ALICE detector:
A. I. Alikhanyan National Science Laboratory (Yerevan Physics Institute) Foundation (ANSL), State Committee of Science and World Federation of Scientists (WFS), Armenia;
Austrian Academy of Sciences, Austrian Science Fund (FWF): [M 2467-N36] and Nationalstiftung f\"{u}r Forschung, Technologie und Entwicklung, Austria;
Ministry of Communications and High Technologies, National Nuclear Research Center, Azerbaijan;
Conselho Nacional de Desenvolvimento Cient\'{\i}fico e Tecnol\'{o}gico (CNPq), Financiadora de Estudos e Projetos (Finep), Funda\c{c}\~{a}o de Amparo \`{a} Pesquisa do Estado de S\~{a}o Paulo (FAPESP) and Universidade Federal do Rio Grande do Sul (UFRGS), Brazil;
Ministry of Education of China (MOEC) , Ministry of Science \& Technology of China (MSTC) and National Natural Science Foundation of China (NSFC), China;
Ministry of Science and Education and Croatian Science Foundation, Croatia;
Centro de Aplicaciones Tecnol\'{o}gicas y Desarrollo Nuclear (CEADEN), Cubaenerg\'{\i}a, Cuba;
Ministry of Education, Youth and Sports of the Czech Republic, Czech Republic;
The Danish Council for Independent Research | Natural Sciences, the VILLUM FONDEN and Danish National Research Foundation (DNRF), Denmark;
Helsinki Institute of Physics (HIP), Finland;
Commissariat \`{a} l'Energie Atomique (CEA), Institut National de Physique Nucl\'{e}aire et de Physique des Particules (IN2P3) and Centre National de la Recherche Scientifique (CNRS) and R\'{e}gion des  Pays de la Loire, France;
Bundesministerium f\"{u}r Bildung und Forschung (BMBF) and GSI Helmholtzzentrum f\"{u}r Schwerionenforschung GmbH, Germany;
General Secretariat for Research and Technology, Ministry of Education, Research and Religions, Greece;
National Research, Development and Innovation Office, Hungary;
Department of Atomic Energy Government of India (DAE), Department of Science and Technology, Government of India (DST), University Grants Commission, Government of India (UGC) and Council of Scientific and Industrial Research (CSIR), India;
Indonesian Institute of Science, Indonesia;
Centro Fermi - Museo Storico della Fisica e Centro Studi e Ricerche Enrico Fermi and Istituto Nazionale di Fisica Nucleare (INFN), Italy;
Institute for Innovative Science and Technology , Nagasaki Institute of Applied Science (IIST), Japanese Ministry of Education, Culture, Sports, Science and Technology (MEXT) and Japan Society for the Promotion of Science (JSPS) KAKENHI, Japan;
Consejo Nacional de Ciencia (CONACYT) y Tecnolog\'{i}a, through Fondo de Cooperaci\'{o}n Internacional en Ciencia y Tecnolog\'{i}a (FONCICYT) and Direcci\'{o}n General de Asuntos del Personal Academico (DGAPA), Mexico;
Nederlandse Organisatie voor Wetenschappelijk Onderzoek (NWO), Netherlands;
The Research Council of Norway, Norway;
Commission on Science and Technology for Sustainable Development in the South (COMSATS), Pakistan;
Pontificia Universidad Cat\'{o}lica del Per\'{u}, Peru;
Ministry of Science and Higher Education and National Science Centre, Poland;
Korea Institute of Science and Technology Information and National Research Foundation of Korea (NRF), Republic of Korea;
Ministry of Education and Scientific Research, Institute of Atomic Physics and Ministry of Research and Innovation and Institute of Atomic Physics, Romania;
Joint Institute for Nuclear Research (JINR), Ministry of Education and Science of the Russian Federation, National Research Centre Kurchatov Institute, Russian Science Foundation and Russian Foundation for Basic Research, Russia;
Ministry of Education, Science, Research and Sport of the Slovak Republic, Slovakia;
National Research Foundation of South Africa, South Africa;
Swedish Research Council (VR) and Knut \& Alice Wallenberg Foundation (KAW), Sweden;
European Organization for Nuclear Research, Switzerland;
Suranaree University of Technology (SUT), National Science and Technology Development Agency (NSDTA) and Office of the Higher Education Commission under NRU project of Thailand, Thailand;
Turkish Atomic Energy Agency (TAEK), Turkey;
National Academy of  Sciences of Ukraine, Ukraine;
Science and Technology Facilities Council (STFC), United Kingdom;
National Science Foundation of the United States of America (NSF) and United States Department of Energy, Office of Nuclear Physics (DOE NP), United States of America.    
\end{acknowledgement}

\bibliographystyle{utphys}   
\bibliography{biblio}

\newpage
\appendix
\section{Appendix}
\label{app:appendix}
\subsection{Charged-particle number density \Nch and $\sum \pt$ distributions with $\ptmin >$ 0.5 and 1.0 $\gevc$ }
\begin{figure*}[htbp]
 \begin{center}
  \includegraphics[width= 0.43\textwidth]{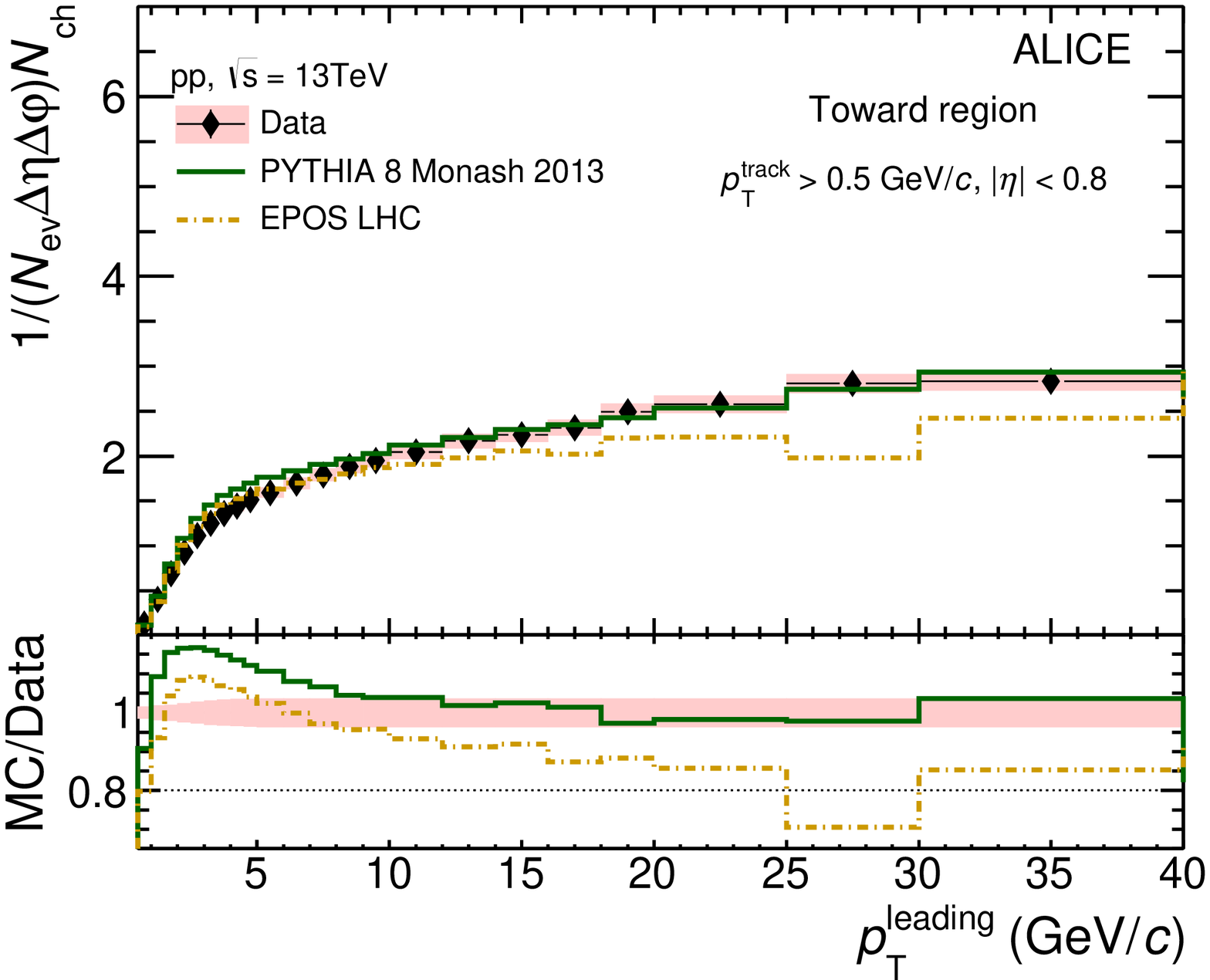}
  \includegraphics[width= 0.43\textwidth]{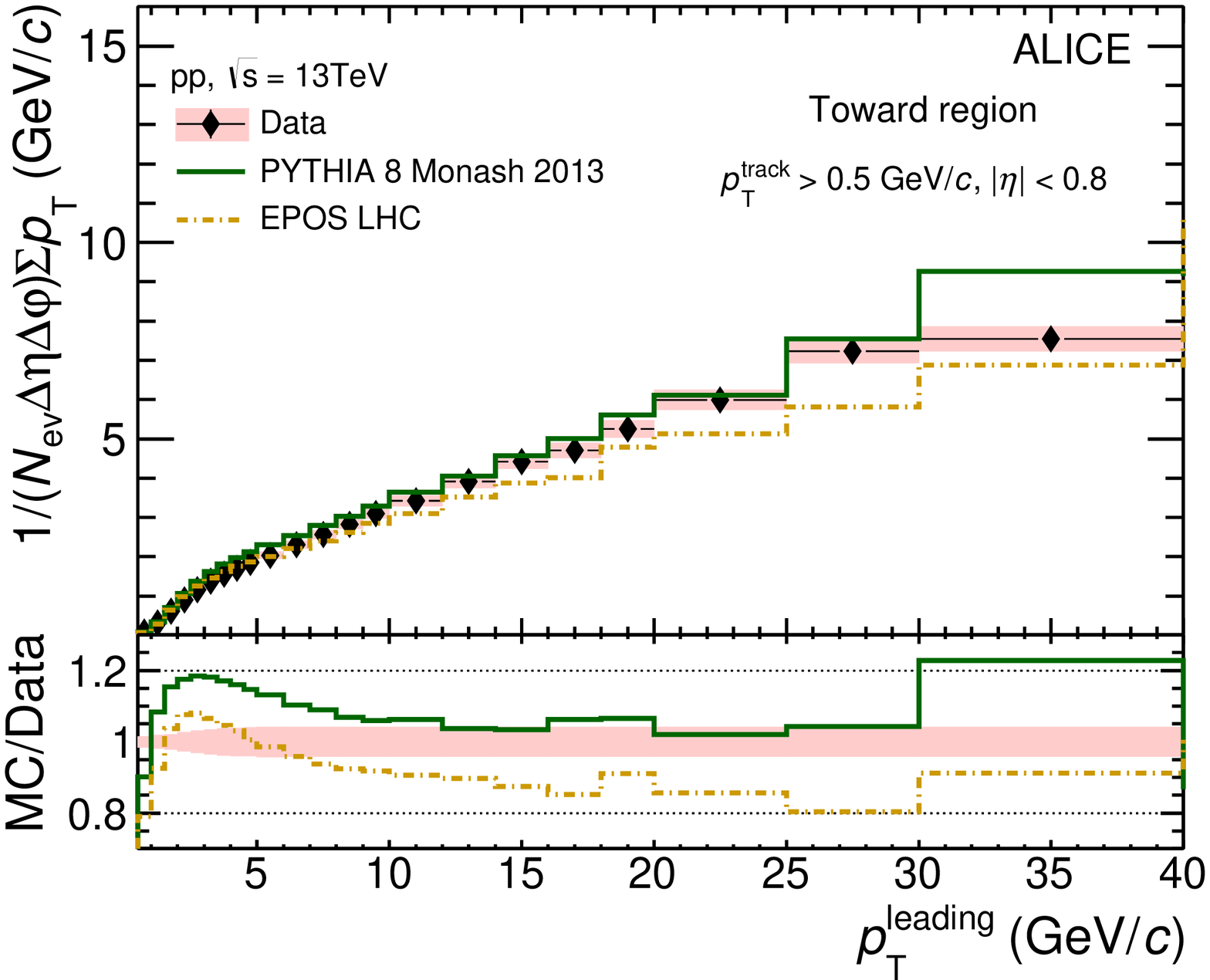}\\
  \includegraphics[width= 0.43\textwidth]{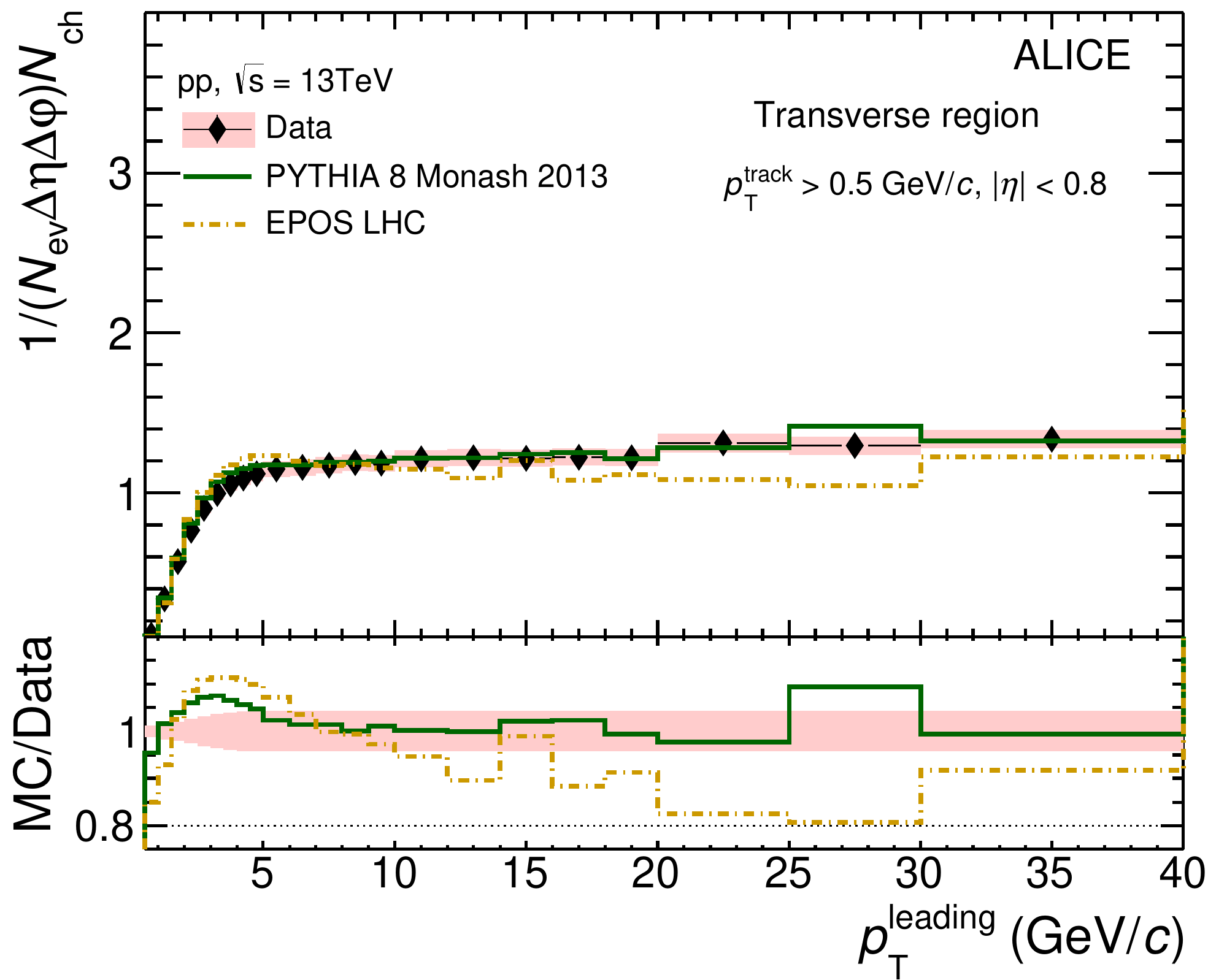}
  \includegraphics[width= 0.43\textwidth]{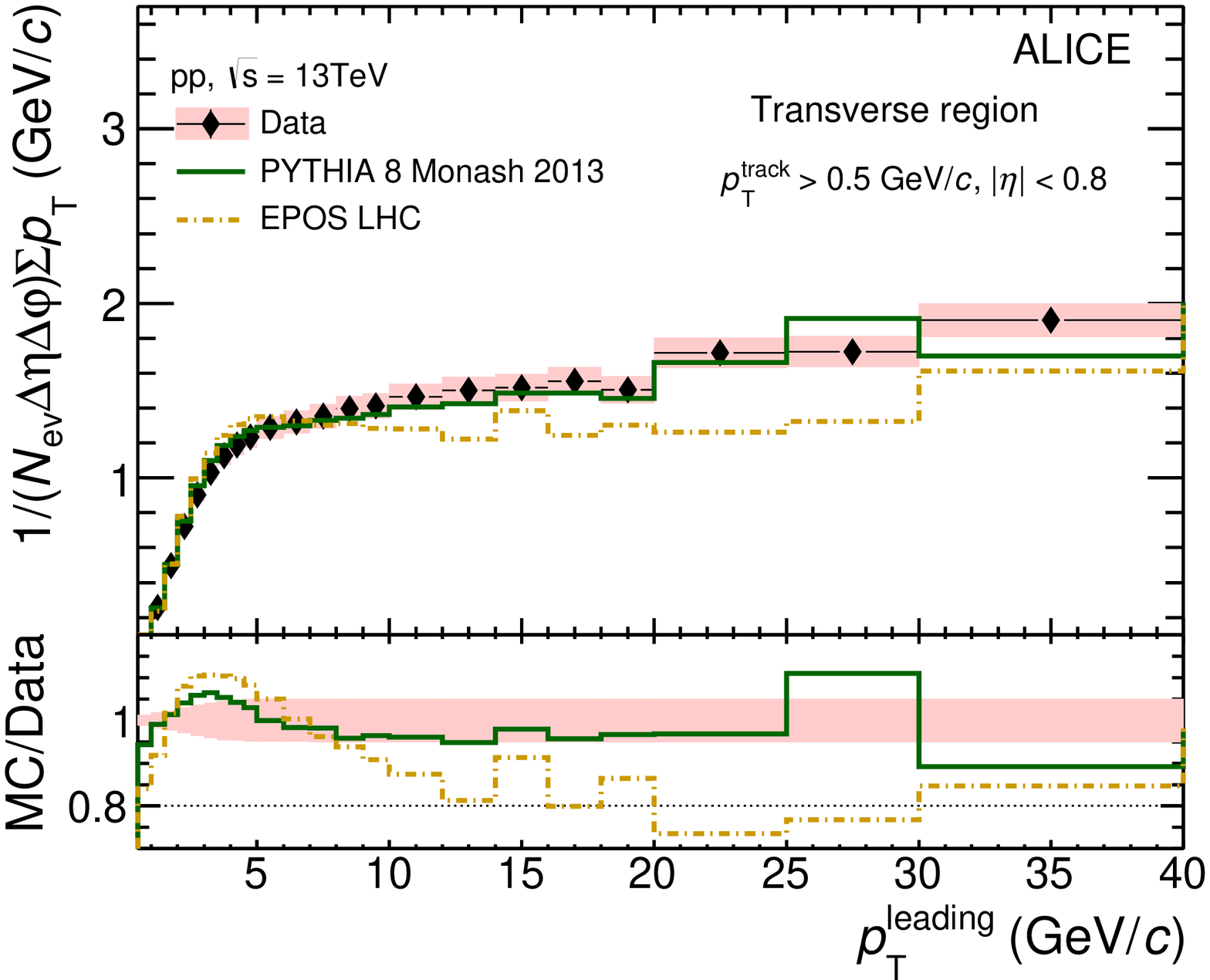}\\
  \includegraphics[width= 0.43\textwidth]{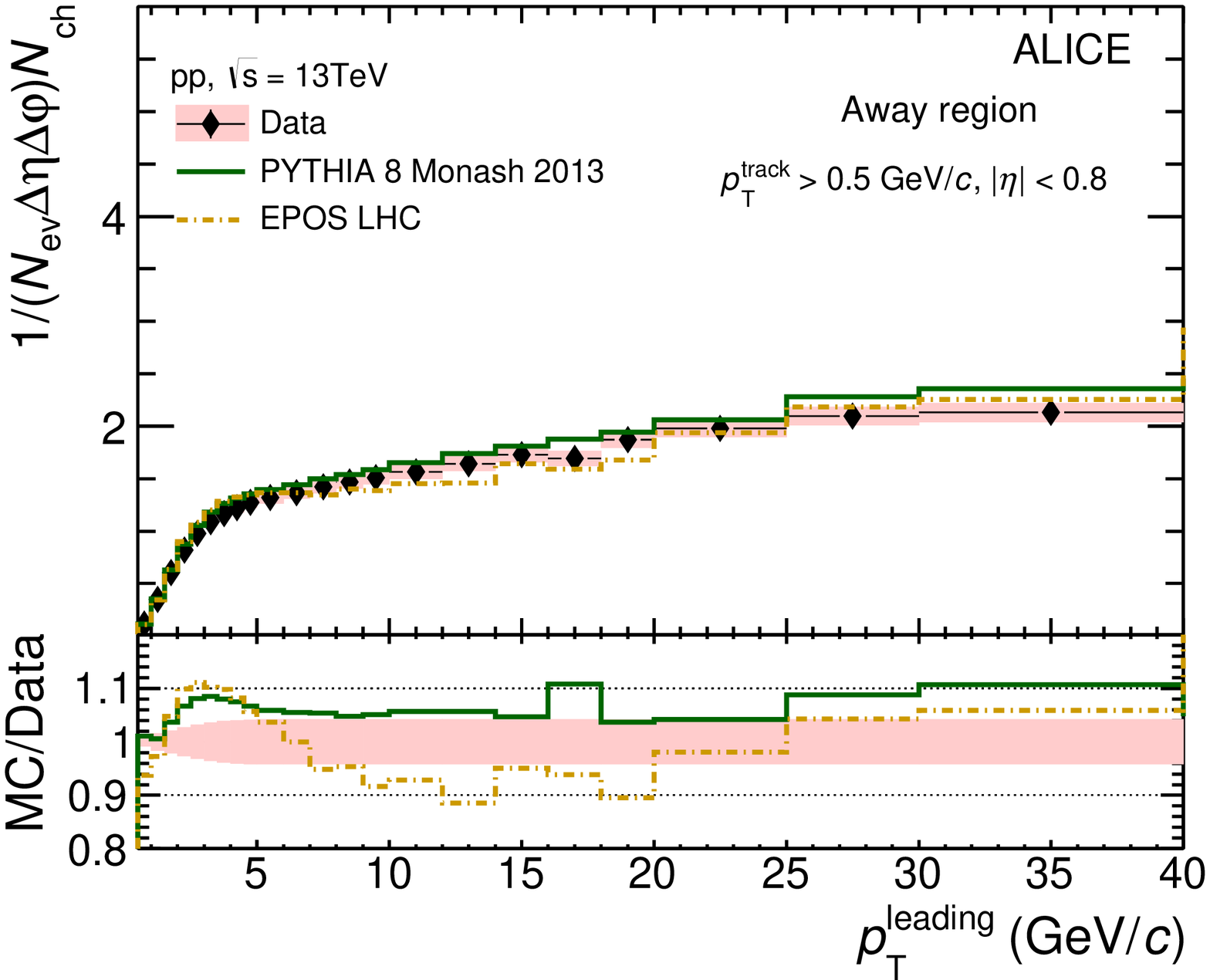}
  \includegraphics[width= 0.43\textwidth]{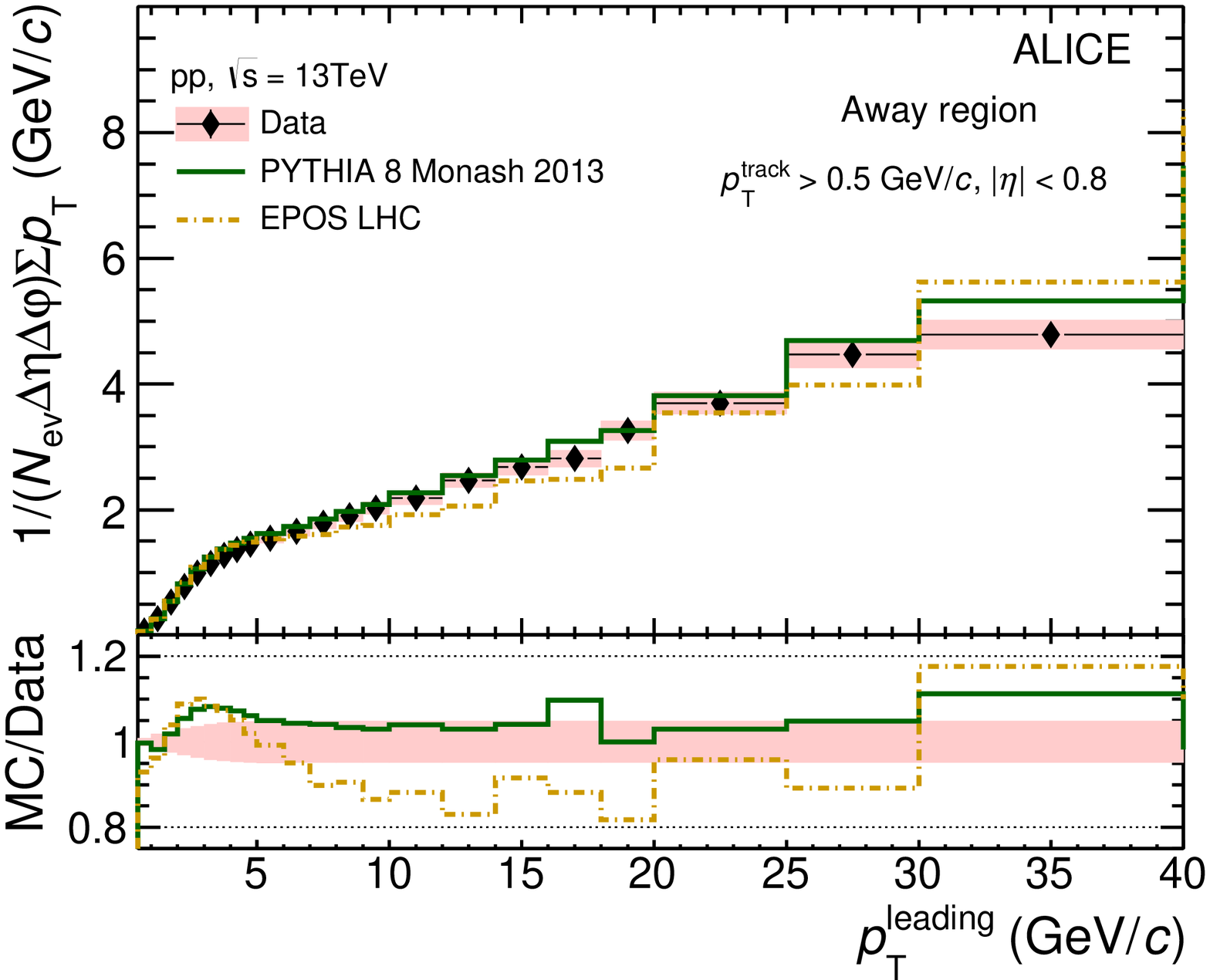}
 \end{center}
 \caption{Number density $\Nch$ (left) and  $\sum \pt$ (right) distributions as a function of \ptlead\, compared to MC predictions in Toward (top), Transverse (middle), and Away (bottom) regions for $\ptmin > 0.5\; \gmom $. The shaded areas in the upper panels represent the systematic uncertainties and vertical error bars indicate statistical uncertainties. In the lower panels, the shaded areas are the sum in quadrature of statistical and systematic uncertainties from the upper panels . No uncertainties are given for the MC calculations.}
 \label{fig: density_05}
\end{figure*}

\begin{figure*}[htbp]
 \begin{center}
  \includegraphics[width= 0.45\textwidth]{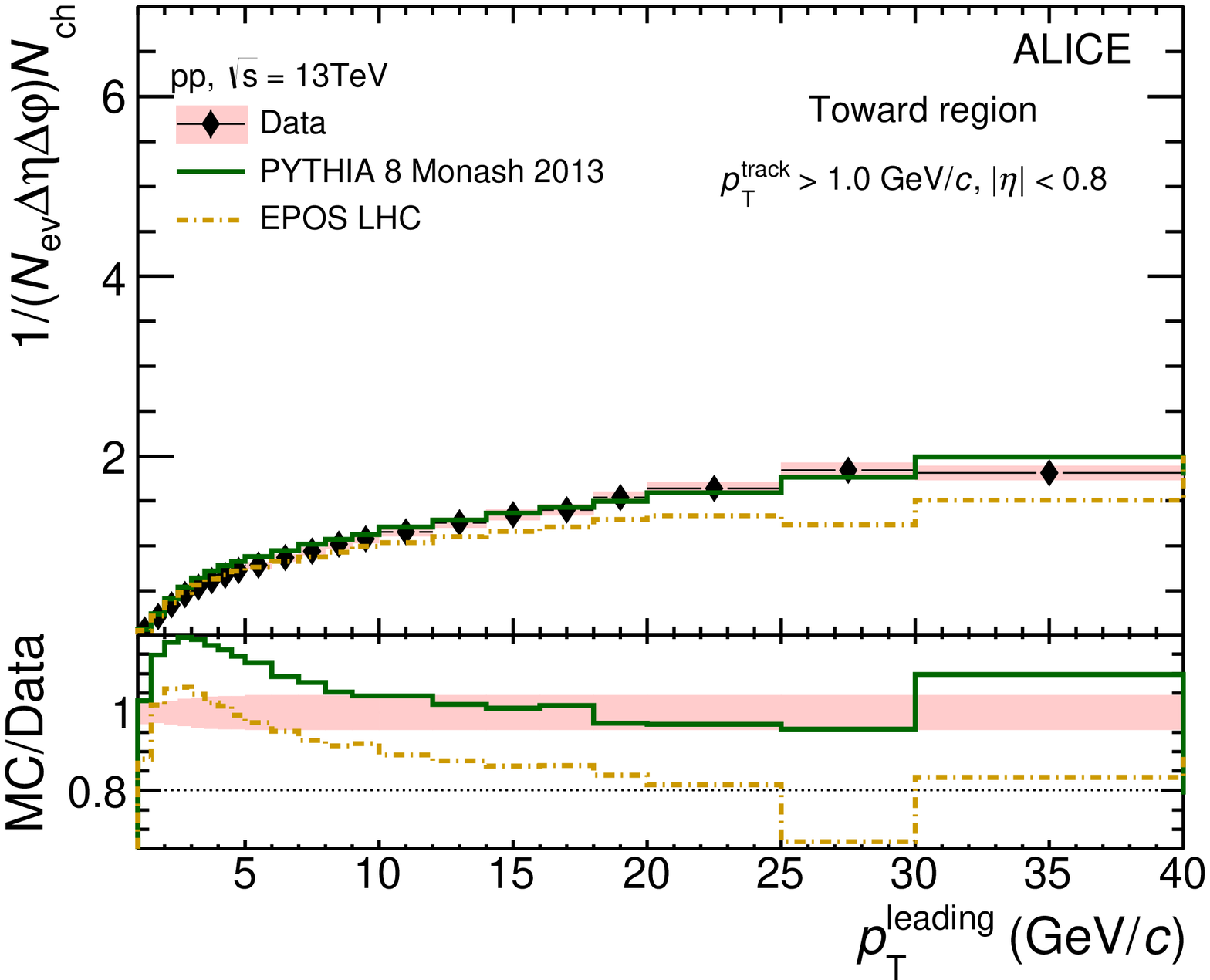}
  \includegraphics[width= 0.45\textwidth]{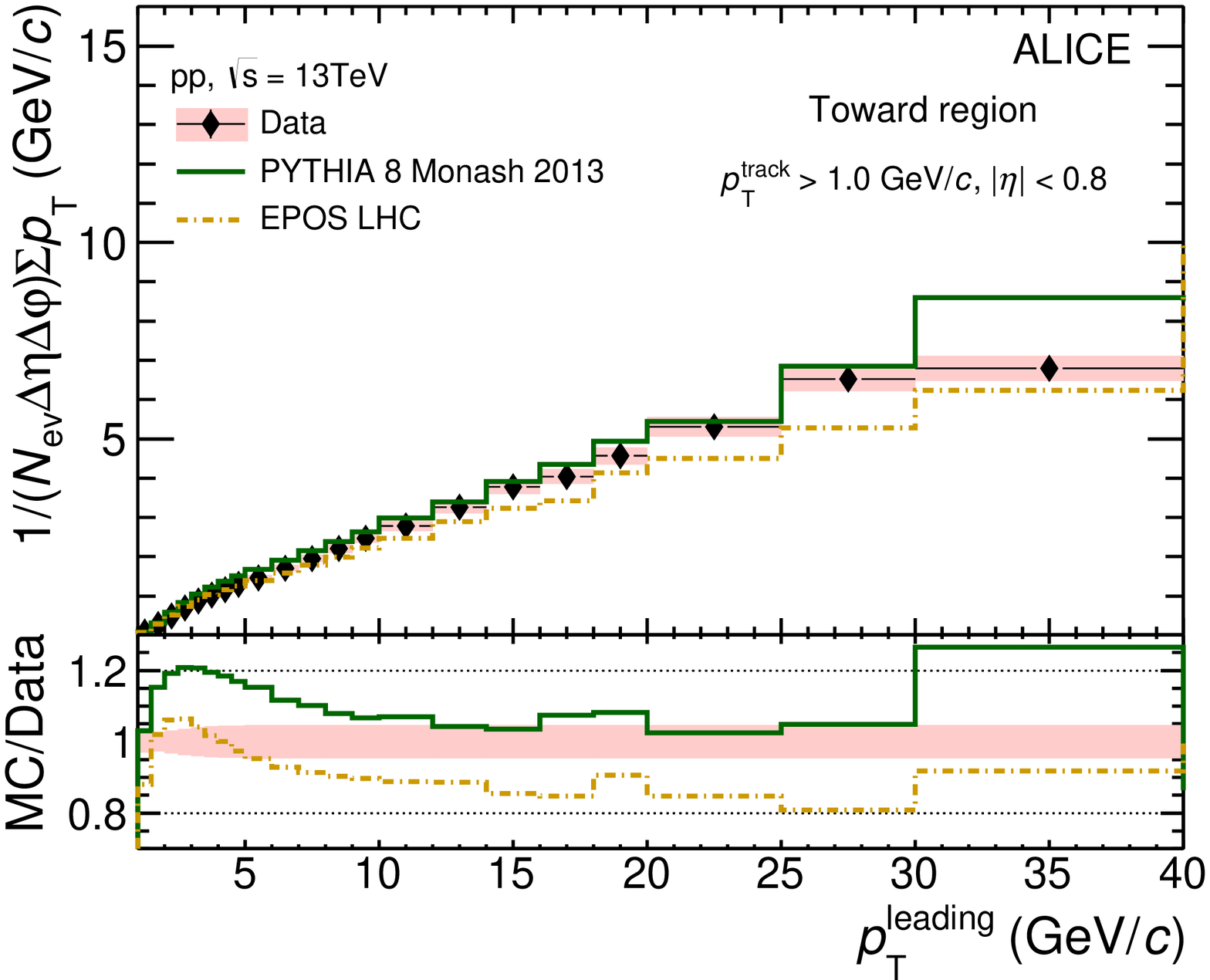}
 \includegraphics[width= 0.45\textwidth]{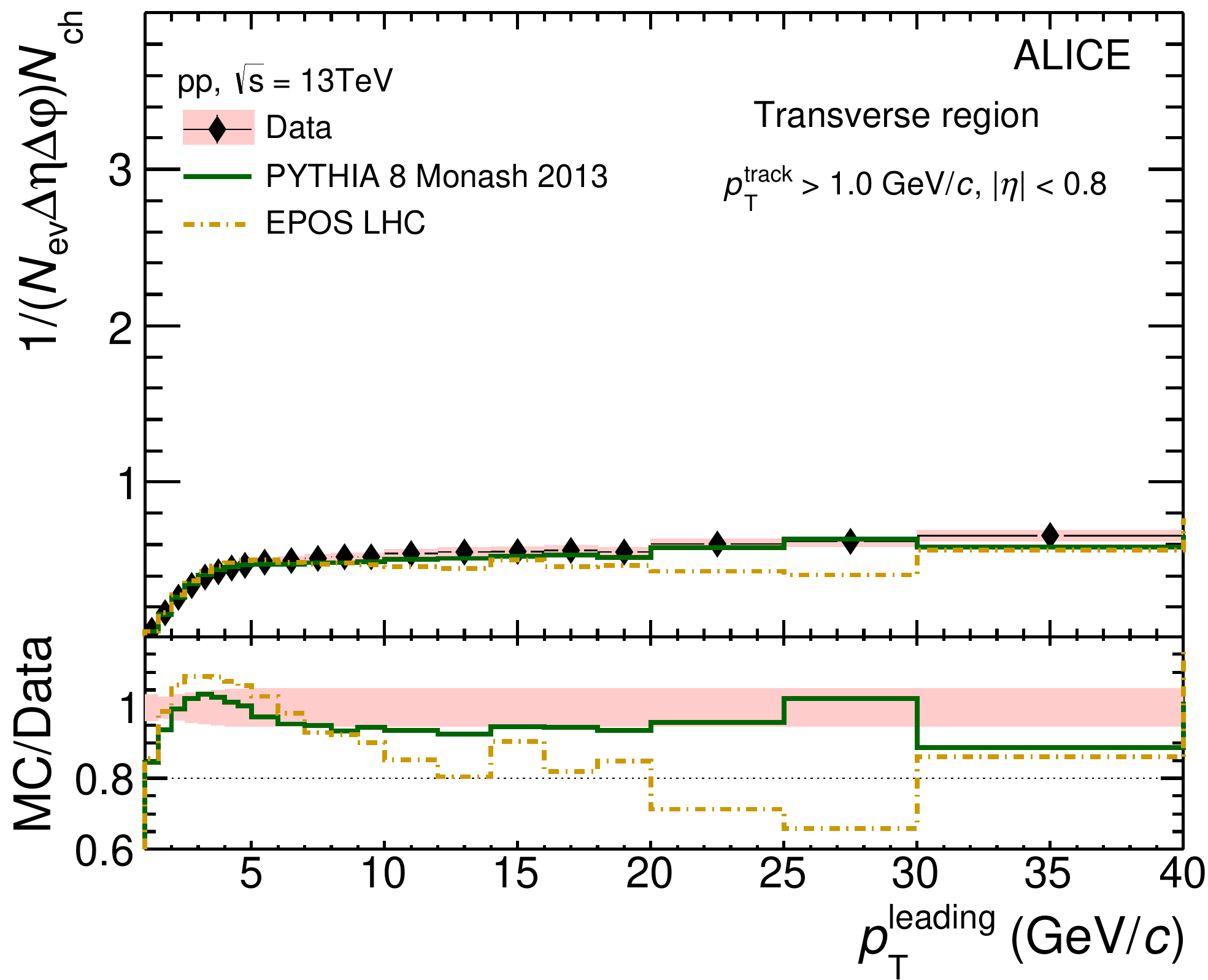}
 \includegraphics[width= 0.45\textwidth]{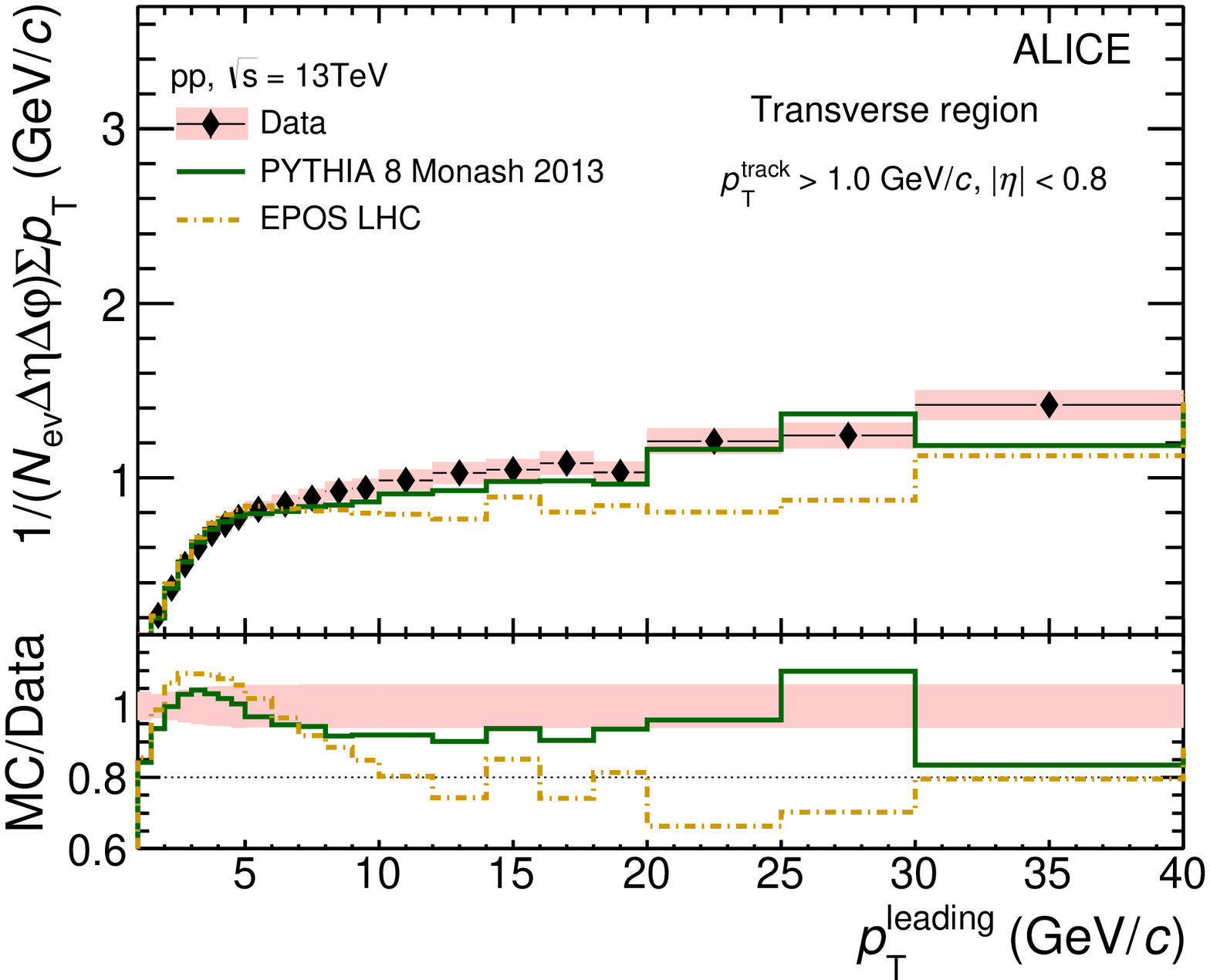}
 \includegraphics[width= 0.45\textwidth]{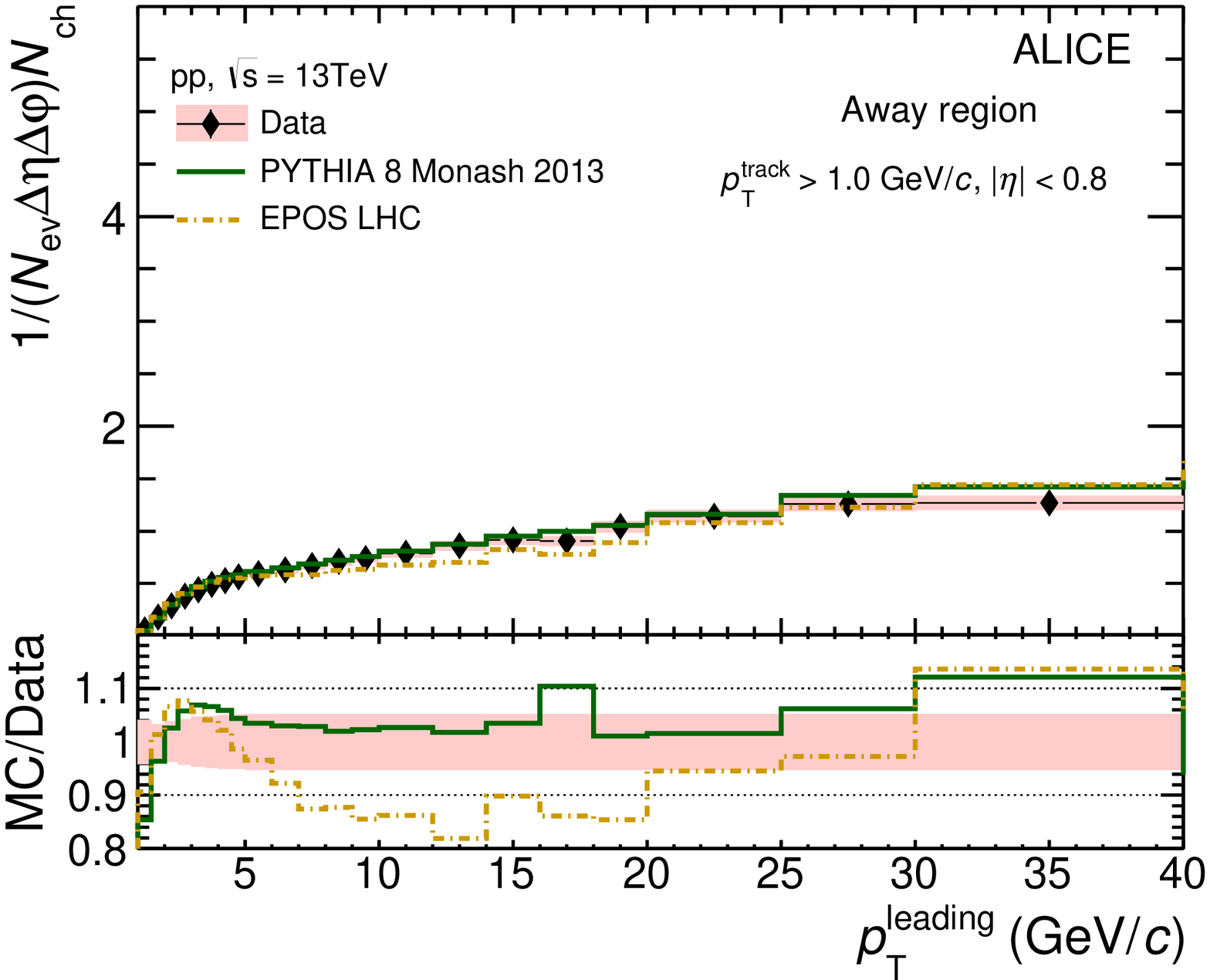}
 \includegraphics[width= 0.45\textwidth]{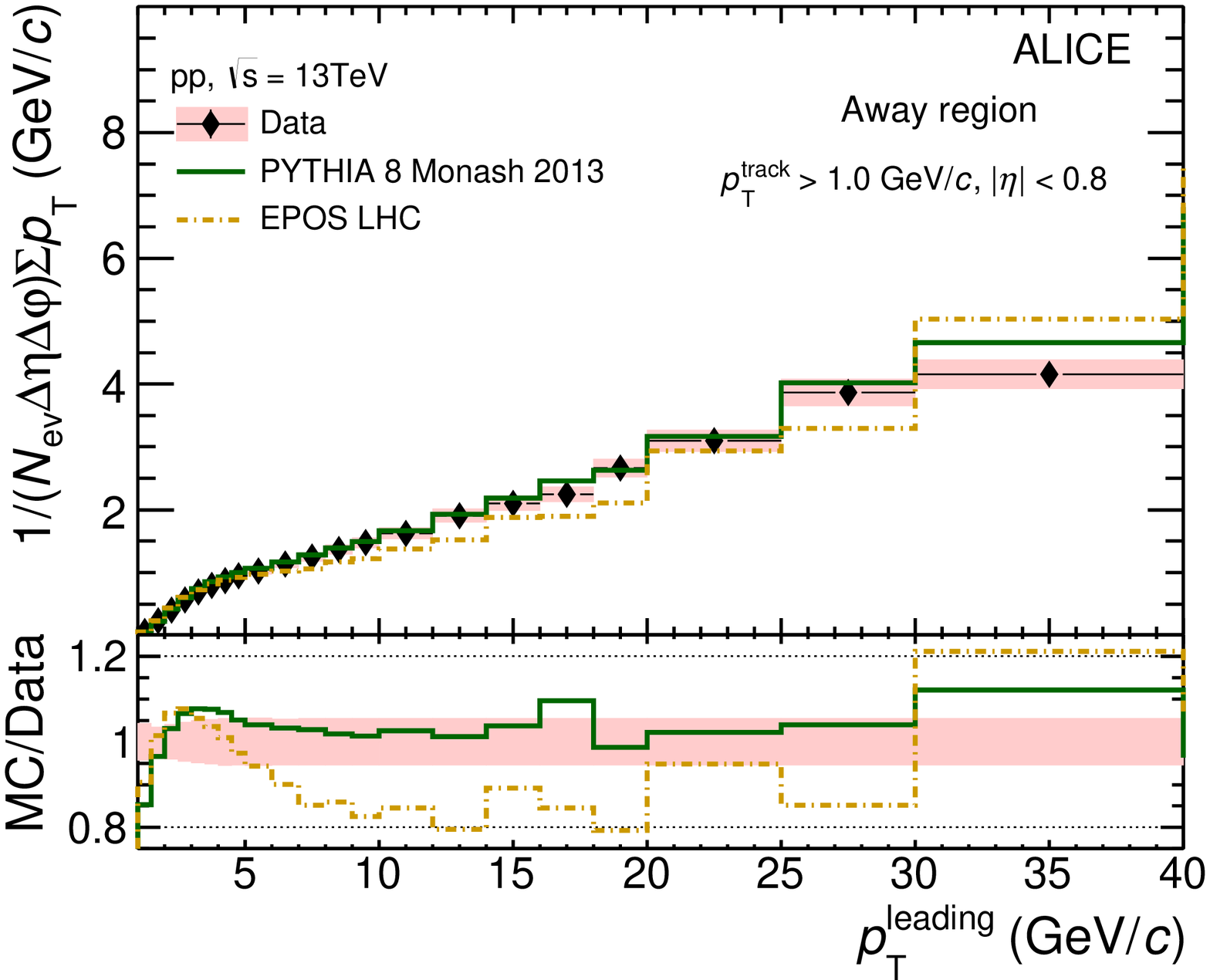}
 \end{center}
 \caption{Number density $\Nch$ (left) and  $\sum \pt$ (right) distributions as a function of \ptlead\ and the comparisons to MC predictions in Toward (top), Transverse (middle), and Away (bottom) regions for $\ptmin > 1.0 \; \gmom $. The shaded areas in the upper panels represent the systematic uncertainties and vertical error bars indicate statistical uncertainties. In the lower panels, the shaded areas are the sum in quadrature of statistical and systematic uncertainties from the upper panels. No uncertainties are given for the MC calculations.}
 \label{fig: density_1}
\end{figure*}

The fully corrected distributions of the averaged charged-particle number and summed-$\pt$ densities as a function of $\ptlead$, in the Toward, Transverse, 
and Away regions for the transverse momentum cuts $\ptmin\ > 0.5$ and $1.0 \; \gmom$ are presented in Fig.~\ref{fig: density_05} and~\ref{fig: density_1}.

\clearpage

\section{The ALICE Collaboration}
\label{app:collab}

\begingroup
\small
\begin{flushleft}
S.~Acharya\Irefn{org141}\And 
D.~Adamov\'{a}\Irefn{org94}\And 
A.~Adler\Irefn{org74}\And 
J.~Adolfsson\Irefn{org80}\And 
M.M.~Aggarwal\Irefn{org99}\And 
G.~Aglieri Rinella\Irefn{org33}\And 
M.~Agnello\Irefn{org30}\And 
N.~Agrawal\Irefn{org10}\textsuperscript{,}\Irefn{org53}\And 
Z.~Ahammed\Irefn{org141}\And 
S.~Ahmad\Irefn{org16}
% ...
\And
\renewcommand\labelenumi{\textsuperscript{\theenumi}~}

\section*{Affiliation notes}
\renewcommand\theenumi{\roman{enumi}}
\begin{Authlist}
\item \Adef{org*}Deceased
\item \Adef{orgI}Dipartimento DET del Politecnico di Torino, Turin, Italy
\item \Adef{orgII}M.V. Lomonosov Moscow State University, D.V. Skobeltsyn Institute of Nuclear, Physics, Moscow, Russia
\item \Adef{orgIII}Department of Applied Physics, Aligarh Muslim University, Aligarh, India
\item \Adef{orgIV}Institute of Theoretical Physics, University of Wroclaw, Poland
\end{Authlist}

\section*{Collaboration Institutes}
\renewcommand\theenumi{\arabic{enumi}~}
\begin{Authlist}
\item \Idef{org1}A.I. Alikhanyan National Science Laboratory (Yerevan Physics Institute) Foundation, Yerevan, Armenia
\item \Idef{org2}Bogolyubov Institute for Theoretical Physics, National Academy of Sciences of Ukraine, Kiev, Ukraine
\item \Idef{org3}Bose Institute, Department of Physics  and Centre for Astroparticle Physics and Space Science (CAPSS), Kolkata, India
\item \Idef{org4}Budker Institute for Nuclear Physics, Novosibirsk, Russia
\item \Idef{org5}California Polytechnic State University, San Luis Obispo, California, United States
\item \Idef{org6}Central China Normal University, Wuhan, China
\item \Idef{org7}Centre de Calcul de l'IN2P3, Villeurbanne, Lyon, France
\item \Idef{org8}Centro de Aplicaciones Tecnol\'{o}gicas y Desarrollo Nuclear (CEADEN), Havana, Cuba
\item \Idef{org9}Centro de Investigaci\'{o}n y de Estudios Avanzados (CINVESTAV), Mexico City and M\'{e}rida, Mexico
\item \Idef{org10}Centro Fermi - Museo Storico della Fisica e Centro Studi e Ricerche ``Enrico Fermi', Rome, Italy
\item \Idef{org11}Chicago State University, Chicago, Illinois, United States
\item \Idef{org12}China Institute of Atomic Energy, Beijing, China
\item \Idef{org13}Comenius University Bratislava, Faculty of Mathematics, Physics and Informatics, Bratislava, Slovakia
\item \Idef{org14}COMSATS University Islamabad, Islamabad, Pakistan
\item \Idef{org15}Creighton University, Omaha, Nebraska, United States
\item \Idef{org16}Department of Physics, Aligarh Muslim University, Aligarh, India
\item \Idef{org17}Department of Physics, Pusan National University, Pusan, Republic of Korea
\item \Idef{org18}Department of Physics, Sejong University, Seoul, Republic of Korea
\item \Idef{org19}Department of Physics, University of California, Berkeley, California, United States
\item \Idef{org20}Department of Physics, University of Oslo, Oslo, Norway
\item \Idef{org21}Department of Physics and Technology, University of Bergen, Bergen, Norway
\item \Idef{org22}Dipartimento di Fisica dell'Universit\`{a} 'La Sapienza' and Sezione INFN, Rome, Italy
\item \Idef{org23}Dipartimento di Fisica dell'Universit\`{a} and Sezione INFN, Cagliari, Italy
\item \Idef{org24}Dipartimento di Fisica dell'Universit\`{a} and Sezione INFN, Trieste, Italy
\item \Idef{org25}Dipartimento di Fisica dell'Universit\`{a} and Sezione INFN, Turin, Italy
\item \Idef{org26}Dipartimento di Fisica e Astronomia dell'Universit\`{a} and Sezione INFN, Bologna, Italy
\item \Idef{org27}Dipartimento di Fisica e Astronomia dell'Universit\`{a} and Sezione INFN, Catania, Italy
\item \Idef{org28}Dipartimento di Fisica e Astronomia dell'Universit\`{a} and Sezione INFN, Padova, Italy
\item \Idef{org29}Dipartimento di Fisica `E.R.~Caianiello' dell'Universit\`{a} and Gruppo Collegato INFN, Salerno, Italy
\item \Idef{org30}Dipartimento DISAT del Politecnico and Sezione INFN, Turin, Italy
\item \Idef{org31}Dipartimento di Scienze e Innovazione Tecnologica dell'Universit\`{a} del Piemonte Orientale and INFN Sezione di Torino, Alessandria, Italy
\item \Idef{org32}Dipartimento Interateneo di Fisica `M.~Merlin' and Sezione INFN, Bari, Italy
\item \Idef{org33}European Organization for Nuclear Research (CERN), Geneva, Switzerland
\item \Idef{org34}Faculty of Electrical Engineering, Mechanical Engineering and Naval Architecture, University of Split, Split, Croatia
\item \Idef{org35}Faculty of Engineering and Science, Western Norway University of Applied Sciences, Bergen, Norway
\item \Idef{org36}Faculty of Nuclear Sciences and Physical Engineering, Czech Technical University in Prague, Prague, Czech Republic
\item \Idef{org37}Faculty of Science, P.J.~\v{S}af\'{a}rik University, Ko\v{s}ice, Slovakia
\item \Idef{org38}Frankfurt Institute for Advanced Studies, Johann Wolfgang Goethe-Universit\"{a}t Frankfurt, Frankfurt, Germany
\item \Idef{org39}Fudan University, Shanghai, China
\item \Idef{org40}Gangneung-Wonju National University, Gangneung, Republic of Korea
\item \Idef{org41}Gauhati University, Department of Physics, Guwahati, India
\item \Idef{org42}Helmholtz-Institut f\"{u}r Strahlen- und Kernphysik, Rheinische Friedrich-Wilhelms-Universit\"{a}t Bonn, Bonn, Germany
\item \Idef{org43}Helsinki Institute of Physics (HIP), Helsinki, Finland
\item \Idef{org44}High Energy Physics Group,  Universidad Aut\'{o}noma de Puebla, Puebla, Mexico
\item \Idef{org45}Hiroshima University, Hiroshima, Japan
\item \Idef{org46}Hochschule Worms, Zentrum  f\"{u}r Technologietransfer und Telekommunikation (ZTT), Worms, Germany
\item \Idef{org47}Horia Hulubei National Institute of Physics and Nuclear Engineering, Bucharest, Romania
\item \Idef{org48}Indian Institute of Technology Bombay (IIT), Mumbai, India
\item \Idef{org49}Indian Institute of Technology Indore, Indore, India
\item \Idef{org50}Indonesian Institute of Sciences, Jakarta, Indonesia
\item \Idef{org51}INFN, Laboratori Nazionali di Frascati, Frascati, Italy
\item \Idef{org52}INFN, Sezione di Bari, Bari, Italy
\item \Idef{org53}INFN, Sezione di Bologna, Bologna, Italy
\item \Idef{org54}INFN, Sezione di Cagliari, Cagliari, Italy
\item \Idef{org55}INFN, Sezione di Catania, Catania, Italy
\item \Idef{org56}INFN, Sezione di Padova, Padova, Italy
\item \Idef{org57}INFN, Sezione di Roma, Rome, Italy
\item \Idef{org58}INFN, Sezione di Torino, Turin, Italy
\item \Idef{org59}INFN, Sezione di Trieste, Trieste, Italy
\item \Idef{org60}Inha University, Incheon, Republic of Korea
\item \Idef{org61}Institut de Physique Nucl\'{e}aire d'Orsay (IPNO), Institut National de Physique Nucl\'{e}aire et de Physique des Particules (IN2P3/CNRS), Universit\'{e} de Paris-Sud, Universit\'{e} Paris-Saclay, Orsay, France
\item \Idef{org62}Institute for Nuclear Research, Academy of Sciences, Moscow, Russia
\item \Idef{org63}Institute for Subatomic Physics, Utrecht University/Nikhef, Utrecht, Netherlands
\item \Idef{org64}Institute of Experimental Physics, Slovak Academy of Sciences, Ko\v{s}ice, Slovakia
\item \Idef{org65}Institute of Physics, Homi Bhabha National Institute, Bhubaneswar, India
\item \Idef{org66}Institute of Physics of the Czech Academy of Sciences, Prague, Czech Republic
\item \Idef{org67}Institute of Space Science (ISS), Bucharest, Romania
\item \Idef{org68}Institut f\"{u}r Kernphysik, Johann Wolfgang Goethe-Universit\"{a}t Frankfurt, Frankfurt, Germany
\item \Idef{org69}Instituto de Ciencias Nucleares, Universidad Nacional Aut\'{o}noma de M\'{e}xico, Mexico City, Mexico
\item \Idef{org70}Instituto de F\'{i}sica, Universidade Federal do Rio Grande do Sul (UFRGS), Porto Alegre, Brazil
\item \Idef{org71}Instituto de F\'{\i}sica, Universidad Nacional Aut\'{o}noma de M\'{e}xico, Mexico City, Mexico
\item \Idef{org72}iThemba LABS, National Research Foundation, Somerset West, South Africa
\item \Idef{org73}Jeonbuk National University, Jeonju, Republic of Korea
\item \Idef{org74}Johann-Wolfgang-Goethe Universit\"{a}t Frankfurt Institut f\"{u}r Informatik, Fachbereich Informatik und Mathematik, Frankfurt, Germany
\item \Idef{org75}Joint Institute for Nuclear Research (JINR), Dubna, Russia
\item \Idef{org76}Korea Institute of Science and Technology Information, Daejeon, Republic of Korea
\item \Idef{org77}KTO Karatay University, Konya, Turkey
\item \Idef{org78}Laboratoire de Physique Subatomique et de Cosmologie, Universit\'{e} Grenoble-Alpes, CNRS-IN2P3, Grenoble, France
\item \Idef{org79}Lawrence Berkeley National Laboratory, Berkeley, California, United States
\item \Idef{org80}Lund University Department of Physics, Division of Particle Physics, Lund, Sweden
\item \Idef{org81}Nagasaki Institute of Applied Science, Nagasaki, Japan
\item \Idef{org82}Nara Women{'}s University (NWU), Nara, Japan
\item \Idef{org83}National and Kapodistrian University of Athens, School of Science, Department of Physics , Athens, Greece
\item \Idef{org84}National Centre for Nuclear Research, Warsaw, Poland
\item \Idef{org85}National Institute of Science Education and Research, Homi Bhabha National Institute, Jatni, India
\item \Idef{org86}National Nuclear Research Center, Baku, Azerbaijan
\item \Idef{org87}National Research Centre Kurchatov Institute, Moscow, Russia
\item \Idef{org88}Niels Bohr Institute, University of Copenhagen, Copenhagen, Denmark
\item \Idef{org89}Nikhef, National institute for subatomic physics, Amsterdam, Netherlands
\item \Idef{org90}NRC Kurchatov Institute IHEP, Protvino, Russia
\item \Idef{org91}NRC «Kurchatov Institute»  - ITEP, Moscow, Russia
\item \Idef{org92}NRNU Moscow Engineering Physics Institute, Moscow, Russia
\item \Idef{org93}Nuclear Physics Group, STFC Daresbury Laboratory, Daresbury, United Kingdom
\item \Idef{org94}Nuclear Physics Institute of the Czech Academy of Sciences, \v{R}e\v{z} u Prahy, Czech Republic
\item \Idef{org95}Oak Ridge National Laboratory, Oak Ridge, Tennessee, United States
\item \Idef{org96}Ohio State University, Columbus, Ohio, United States
\item \Idef{org97}Petersburg Nuclear Physics Institute, Gatchina, Russia
\item \Idef{org98}Physics department, Faculty of science, University of Zagreb, Zagreb, Croatia
\item \Idef{org99}Physics Department, Panjab University, Chandigarh, India
\item \Idef{org100}Physics Department, University of Jammu, Jammu, India
\item \Idef{org101}Physics Department, University of Rajasthan, Jaipur, India
\item \Idef{org102}Physikalisches Institut, Eberhard-Karls-Universit\"{a}t T\"{u}bingen, T\"{u}bingen, Germany
\item \Idef{org103}Physikalisches Institut, Ruprecht-Karls-Universit\"{a}t Heidelberg, Heidelberg, Germany
\item \Idef{org104}Physik Department, Technische Universit\"{a}t M\"{u}nchen, Munich, Germany
\item \Idef{org105}Politecnico di Bari, Bari, Italy
\item \Idef{org106}Research Division and ExtreMe Matter Institute EMMI, GSI Helmholtzzentrum f\"ur Schwerionenforschung GmbH, Darmstadt, Germany
\item \Idef{org107}Rudjer Bo\v{s}kovi\'{c} Institute, Zagreb, Croatia
\item \Idef{org108}Russian Federal Nuclear Center (VNIIEF), Sarov, Russia
\item \Idef{org109}Saha Institute of Nuclear Physics, Homi Bhabha National Institute, Kolkata, India
\item \Idef{org110}School of Physics and Astronomy, University of Birmingham, Birmingham, United Kingdom
\item \Idef{org111}Secci\'{o}n F\'{\i}sica, Departamento de Ciencias, Pontificia Universidad Cat\'{o}lica del Per\'{u}, Lima, Peru
\item \Idef{org112}St. Petersburg State University, St. Petersburg, Russia
\item \Idef{org113}Stefan Meyer Institut f\"{u}r Subatomare Physik (SMI), Vienna, Austria
\item \Idef{org114}SUBATECH, IMT Atlantique, Universit\'{e} de Nantes, CNRS-IN2P3, Nantes, France
\item \Idef{org115}Suranaree University of Technology, Nakhon Ratchasima, Thailand
\item \Idef{org116}Technical University of Ko\v{s}ice, Ko\v{s}ice, Slovakia
\item \Idef{org117}Technische Universit\"{a}t M\"{u}nchen, Excellence Cluster 'Universe', Munich, Germany
\item \Idef{org118}The Henryk Niewodniczanski Institute of Nuclear Physics, Polish Academy of Sciences, Cracow, Poland
\item \Idef{org119}The University of Texas at Austin, Austin, Texas, United States
\item \Idef{org120}Universidad Aut\'{o}noma de Sinaloa, Culiac\'{a}n, Mexico
\item \Idef{org121}Universidade de S\~{a}o Paulo (USP), S\~{a}o Paulo, Brazil
\item \Idef{org122}Universidade Estadual de Campinas (UNICAMP), Campinas, Brazil
\item \Idef{org123}Universidade Federal do ABC, Santo Andre, Brazil
\item \Idef{org124}University of Cape Town, Cape Town, South Africa
\item \Idef{org125}University of Houston, Houston, Texas, United States
\item \Idef{org126}University of Jyv\"{a}skyl\"{a}, Jyv\"{a}skyl\"{a}, Finland
\item \Idef{org127}University of Liverpool, Liverpool, United Kingdom
\item \Idef{org128}University of Science and Technology of China, Hefei, China
\item \Idef{org129}University of South-Eastern Norway, Tonsberg, Norway
\item \Idef{org130}University of Tennessee, Knoxville, Tennessee, United States
\item \Idef{org131}University of the Witwatersrand, Johannesburg, South Africa
\item \Idef{org132}University of Tokyo, Tokyo, Japan
\item \Idef{org133}University of Tsukuba, Tsukuba, Japan
\item \Idef{org134}Universit\'{e} Clermont Auvergne, CNRS/IN2P3, LPC, Clermont-Ferrand, France
\item \Idef{org135}Universit\'{e} de Lyon, Universit\'{e} Lyon 1, CNRS/IN2P3, IPN-Lyon, Villeurbanne, Lyon, France
\item \Idef{org136}Universit\'{e} de Strasbourg, CNRS, IPHC UMR 7178, F-67000 Strasbourg, France, Strasbourg, France
\item \Idef{org137}Universit\'{e} Paris-Saclay Centre d'Etudes de Saclay (CEA), IRFU, D\'{e}partment de Physique Nucl\'{e}aire (DPhN), Saclay, France
\item \Idef{org138}Universit\`{a} degli Studi di Foggia, Foggia, Italy
\item \Idef{org139}Universit\`{a} degli Studi di Pavia, Pavia, Italy
\item \Idef{org140}Universit\`{a} di Brescia, Brescia, Italy
\item \Idef{org141}Variable Energy Cyclotron Centre, Homi Bhabha National Institute, Kolkata, India
\item \Idef{org142}Warsaw University of Technology, Warsaw, Poland
\item \Idef{org143}Wayne State University, Detroit, Michigan, United States
\item \Idef{org144}Westf\"{a}lische Wilhelms-Universit\"{a}t M\"{u}nster, Institut f\"{u}r Kernphysik, M\"{u}nster, Germany
\item \Idef{org145}Wigner Research Centre for Physics, Budapest, Hungary
\item \Idef{org146}Yale University, New Haven, Connecticut, United States
\item \Idef{org147}Yonsei University, Seoul, Republic of Korea
\end{Authlist}
\endgroup
\end{document}